\documentclass[%
 reprint,
 amsmath,amssymb,
 aps,
floatfix,
]{revtex4-1}
\usepackage{graphicx}
\usepackage{dcolumn}
\usepackage{bm}
\usepackage{color}
\usepackage[utf8]{inputenc}
\usepackage{capt-of}
\usepackage{tabu}
\usepackage{mwe} 
\usepackage{subcaption}
\usepackage{soul}
\usepackage{xr}

\makeatletter
\newcommand*{\addFileDependency}[1]{
  \typeout{(#1)}
  \@addtofilelist{#1}
  \IfFileExists{#1}{}{\typeout{No file #1.}}
}
\makeatother

\newcommand*{\myexternaldocument}[1]{%
    \externaldocument{#1}%
    \addFileDependency{#1.tex}%
    \addFileDependency{#1.aux}%
}

\myexternaldocument{sm}

\begin{document}

\title{Phases and homogeneous ordered states in alignment-based \\ self-propelled particle models}

\author{Yinong Zhao$^{1,2}$, Thomas Ihle$^{3}$, Zhangang Han$^{4}$, Cristi\'an Huepe$^{4,5,6}$ and Pawel Romanczuk$^{1,2}$}

\address{$^1$Institute for Theoretical Biology, Department of Biology, Humboldt-Universität zu Berlin, 10115 Berlin, Germany}
\address{$^2$Bernstein Center for Computational Neuroscience Berlin, 10115 Berlin, Germany}
\address{$^3$ Insitute of Physics, University of Greifswald, 17489 Greifswald, Germany}
\address{$^4$School of Systems Science, Beijing Normal University, Beijing 100875, China}
\address{$^5$CHuepe Labs, 2713 W Haddon Ave \#1, Chicago, IL 60622, USA}
\address{$^6$Northwestern Institute on Complex Systems and ESAM, Northwestern University, Evanston, IL 60208, USA}

\begin{abstract}
We study a set of models of self-propelled particles that achieve collective motion through similar alignment-based dynamics, considering versions with and without repulsive interactions that do not affect the heading directions.
We explore their phase space within a broad range of values of two nondimensional parameters (coupling strength and Peclet number), characterizing their polarization and degree of clustering.
The resulting phase diagrams display equivalent, similarly distributed regions for all models with repulsion.
The diagrams without repulsion exhibit differences, in particular for high coupling strengths.
We compare the boundaries and representative states of all regions, identifying various regimes that had not been previously characterized.
We analyze in detail three types of homogeneous polarized states, comparing them to existing theoretical and numerical results by computing their velocity and density correlations, giant number fluctuations, and local order-density coupling. We find that they all deviate in one way or another from the theoretical predictions, attributing these differences either to the remaining inhomogeneities or to finite-size effects.
We discuss our results in terms of the universal or specific features of each model, their thermodynamic limit, and the high mixing and low mixing regimes.
Our study provides a broad, overarching perspective on the multiple phases and states found in alignment-based self-propelled particle models.
\end{abstract}

\maketitle

\section{Introduction}


The spontaneous emergence of collective dynamics in groups of active, self-propelled components is widely observed in nature. 
A variety of animals, including insects \cite{theraulaz2002spatial, bazazi2011nutritional,DeThViReview2012}, birds \cite{cavagna2010scale,cavagna2015short,ballerini2008interaction}, fish \cite{Breder_54, couzin05,DeThViReview2012}, and mammals \cite{westley2018collective}, can achieve long-range ordered movement through short-range interactions.
In addition, groups of non-living self-propelled components have also been shown to exhibit self-organized collective motion \cite{deseigne2010collective,geyer2018sounds,lavergne2019group}.
The analysis of these systems has led to multiple fundamental questions, across disciplines.
What emergent structures and dynamics can be observed?
How are these connected to individual motion and interaction rules?
Are there universal coarse-grained states that are independent of specific details?

Self-propelled particle (SPP) models have been an important tool for exploring the general collective properties of active systems, without focusing on specific living or nonliving agents \cite{SumpterBook, DeThViReview2012, romanczuk2012active,VicsekZafeirisReview2012, SwarmRoboticsReview2013}.
In these models, the active components are represented by particles that self-propel along their heading directions, with velocities that depend on the states and dynamics of other particles, such as the relative positions or velocities of neighbors.
Many of the commonly used interaction rules are based on a tendency to (ferromagnetically) align with neighboring particles, since this can lead to large-scale collective motion by directly matching the local individual velocities or headings.


One of the simplest possible and most widely used SPP models is the Vicsek model \cite{vicsek1995novel}. This is an alignment-based algorithm that was introduced as a non-equilibrium extension of the classical XY model.
Here, spins are replaced by particles that advance at a fixed speed in their pointing direction and instantaneously align at each timestep to the average heading of all particles within a given interaction range.
The control parameters of the standard Vicsek model can be reduced to the scaled mean density of the system, the intensity of an added orientational noise with respect to the velocity, and the ratio of the mean-free path to the radius of interaction.
Despite its simplicity, the Vicsek model can produce a surprising variety of ordered and disordered collective states with different density distributions.
These have attracted significant attention over the past decades, such as those involving the emergence of long-range orientational order, clustering, moving bands and cross-sea patterns \cite{barberis2018emergence,gregoire2004onset,kursten2020dry}.


A number of more realistic SPP algorithms have extended the Vicsek model in different ways to consider, for example, the gradual alignment of particles in continuous time, repulsive interactions, and speed-density or speed-order coupling 
\cite{morin2015collective,martin2018collective,czirok1996formation,mishra2012collective}.
Many studies have also focused on specific collective states, such as homogeneous order  \cite{toner2005hydrodynamics,chate2008collective,mahault2019quantitative}, bands \cite{gregoire2004onset,solon2015pattern,caussin2014emergent},  motility-induced phase separation \cite{sese2018velocity}, and clustering \cite{barberis2018emergence,martin2018collective, kyriakopoulos2019clustering}.
This has led to a plethora of results that may or may not be universal or specific to given algorithms or simulation choices.
Some research has tried to overcome this issue by using dimensionless quantities that are expected to be model-independent.
In \cite{martin2018collective}, for example, the phase space of an SPP model with continuous alignment and repulsion was explored in terms of two nondimensional parameters that are also valid for other models; the alignment coupling strength and the Peclet number.
Very few works have compared the results of different SPP models, however, to examine which features may be common to all of them and which are model-dependent \cite{HuepeAldanaPhysA2008}.


A different approach for studying universal emerging states in SPP systems has been to derive hydrodynamic field equations, which capture coarse-grained dynamics that are expected to be independent of the microscopic details.
In their seminal work, Toner and Tu first wrote this type of equations using symmetry arguments \cite{ToTuPRE98,toner2005hydrodynamics,toner2012reanalysis}.
Other groups have since derived similar macroscopic descriptions from the microscopic, individual particle level through a variety of methods
\cite{bertin2009hydrodynamic,ihle2011kinetic,grossmann2013self, peshkov2014boltzmann}.
This approach has helped unveil various general features of the SPP systems, their order-disorder transition, and the density-order feedback that leads to spatial structures and fluctuations
\cite{bertin2009hydrodynamic,ihle2011kinetic,grossmann2013self, peshkov2014boltzmann,chate2008collective}. 
It has also been extensively used to compute the analytical properties of perturbations about the homogeneous ordered `flocking' state \cite{ToTuPRE98,toner2005hydrodynamics,toner2012reanalysis}.
These have been tested numerically through SPP simulations, which have partially confirmed their predictions \cite{geyer2018sounds, mahault2019quantitative}.


Despite the multiple numerical and analytical studies of SPP systems that have been carried out for over two decades, many questions still remain open.
One particularly lacking aspect is our understanding of the universality or specificity of the various states that are found using different models and in different regions of the phase space, as well as their connection to the analytical results obtained from the hydrodynamic theory.


In this paper, we address these issues by simulating four different SPP models with similar alignment-based dynamics, with and without repulsive interactions that do not affect the particle headings.
We compare their results for a broad range of parameters, following the approach developed in \cite{martin2018collective}, to explore their phase spaces as a function of the coupling strength and Peclet number. 
We find that all models with repulsion display similar phase diagrams (albeit with distinct quantitative differences) and that the cases without repulsion are only similar for low coupling strengths.
We then focus on three homogeneous aligned states that we identify for different parameter combinations. We measure their velocity and density correlations, giant number fluctuations, and local density-order coupling, comparing their properties to previous numerical studies and to analytical predictions derived from the Toner-Tu theory.


The paper is organized as follows. 
In Section \ref{sec:models}, we use a common framework to introduce all the models that we will study.
In Section \ref{sec:tools}, we describe the tools that we used to explore their parameter space and characterize their collective states. 
Section \ref{sec:overview} provides a detailed overview of the phase space and emergent stationary states of one of these models.
In Section \ref{sec:analysis}, we analyze the statistical properties of the three different homogeneous polarized states identified in this model.
Section \ref{sec:comparison} compares the phase diagrams of all the considered models with repulsion and presents a case without repulsion.
Finally, we end the paper with our discussion and conclusions in Section \ref{sec:conclusion}.

\section{Alignment Models}
\label{sec:models}

We will consider four similar SPP models with continuous time in two spatial dimensions to study how their macroscopic stationary states are affected by the details of their interaction rules.
All these models achieve collective motion through interactions that tend to align nearby neighbors and are defined by dynamical rules that evolve the particle positions and velocities continuously in time.

We formulate first a general overdamped equation of motion in 2D, valid for all models, which reads
\begin{equation}
\label{eq1}
    \dot{\vec{r}}_i = v_0 \hat{n}_i(\theta_i) + \vec{F}_i,
\end{equation}
where $\vec{r}_i$ is the position of particle $i$.
The self-propulsion speed $v_0$ is constant and equal for all particles and the unit vector $\hat{n}_i(\theta_i)=[\cos(\theta_i),\sin(\theta_i)]^T$ points in the heading direction of particle $i$, with angle $\theta_i$.
The displacement force $\vec{F}_i$ acts on the particle position, but not on its heading direction.

In the continuous-time case considered here, we assume the following general equation for the orientation dynamics
\begin{eqnarray}
\label{eq:ThetaDot}
\dot{\theta}_i &=& \frac{1}{\tau}\Omega \left(\theta_i, \{ \theta_{j} \}_{j \in S_i} \right) + \sigma_{\theta} \, \xi_{\theta} \mbox{.}
\end{eqnarray}
Here, $\Omega(\theta_i,\left\{\theta_j\right\}_{j\in S_i})$
is a function that will align, with characteristic time $\tau$, the orientation $\theta_i$ of focal particle $i$ to some type of (effective) average heading direction of its neighbors $j$.
The set of particles $S_i$ interacting with the focal particle $i$ contains all particles within a distance $R$ of $\vec{r}_i$, which corresponds to \emph{metric interactions}, as in the original Vicsek model.
We note that other definitions of neighbors have also been considered, such as selecting a fixed number of the nearest neighbors or the first Voronoi neighbors.
The last term of Eq.~\ref{eq:ThetaDot} adds a $\delta$-correlated Gaussian white noise of strength $\sigma_{\theta}$ through a random variable $\xi_{\theta}$ that satisfies $\langle \xi_{\theta}(t_1)\xi_{\theta}(t_2) \rangle = \delta(t_2 - t_1)$.

The function $\Omega$ can be defined in multiple ways. 
We will implement here the four options detailed below, all of which are based on, or inspired by, previous models in the literature \cite{morin2015collective,martin2018collective,czirok1996formation}.

For the first model, it is natural to consider an algorithm in which the turning forces are proportional to the mean difference of heading angles between interacting particles.
Our \emph{Mean-Angle} (MA) algorithm is thus defined by
\begin{equation}
    \label{eq:model-average-angle}
    \Omega_{MA} = \left\langle \mathrm{mod^*}(\theta_j - \theta_i)\right \rangle_{j\in S_i},
\end{equation}
where $\mathrm{mod}^*(\theta)=\mathrm{mod}(\theta+\pi,2\pi)-\pi$ is a modified modulo function that computes the smallest angular difference between $\theta_j$ and $\theta_i$. Here, we use the following short notation for the average of an arbitrary function $f$ with respect to neighbors within set $S_i$:  
$$\langle f(\vec{v}_i,\vec{v}_j) \rangle_{j\in S_i}=\frac{1}{N_i}\sum_{j\in S_i} f(\vec{v}_i,\vec{v}_j)\ ,$$
with $N_i=\sum_{j\in S_i} 1$.

For our second model, we follow the perspective of a physical process where the alignment force is proportional to the mean differences of velocity vectors $\langle \vec{v}_j -\vec{v}_i \rangle_{j \in S_i}$
\cite{grossmann2012active,morin2015collective}.
For constant speeds and expressing this interaction in terms of angular differences (polar coordinates), we obtain our \emph{Mean-Sine} (MS) model, defined by
\begin{equation}
    \label{eq:model-average-sine}
    \Omega_{MS} = \left\langle\sin(\theta_j -\theta_i)\right\rangle_{j\in S_i}.
\end{equation}
An advantage of $\Omega_{MS}$ is that it has a clearer physical interpretation and continuous derivative. We expect AS and MA interactions to be very similar when agents are highly aligned and we will thus focus on differences between stationary states in less ordered regimes.

Our third model is a variation of the MS case where the mean is replaced by the sum of the sine of the heading angle differences. 
The resulting \emph{Additive-Sine} (AS) model has been considered in the literature and therefore provides a point of comparison \cite{farrell2012pattern,martin2018collective}. 
It is defined by
\begin{equation}
    \label{eq:model-sum-sine}
    \Omega_{AS} = \sum_{j\in S_i}\sin(\theta_j -\theta_i).
\end{equation}
We note that this model will strengthen the effect of the local density, since the alignment force is not divided by the number of neighbors. Some aspects of the difference between the AS and the MS model were recently discussed in \cite{chepizhko2021revisiting, kursten2021quantitative}.

Our fourth and final model is a continuous-time generalization of the Vicsek algorithm. In this \emph{Sine-Velocity} (SV) model, the turning force is proportional to the sine of the angular difference between the heading of the focal particle and an average heading of the neighboring particles. It is thus defined by
\begin{eqnarray}
    \label{eq:model-sine-average}
    \Omega_{SV} &=& \sin(\bar{\theta}_j -\theta_i) \\
    \bar{\theta}_j &=& \mathrm{Angle}\left( \sum_{j\in S_i}\hat{n}(\theta_j) \right).
\end{eqnarray}
Here, $\mathrm{Angle}( \cdot )$ is a function that yields the angle of its argument and $\bar{\theta}_j$ is a Vicsek-style average heading angle.
A similar model was introduced in \cite{czirok1996formation}. 

Finally, we define $\vec{F}_i$ in Eq.~(\ref{eq1}) as a displacement repulsion force that will reduce particle overlap and limit the formation of high density clusters.
In order to decouple its effect from the alignment dynamics, we define $\vec{F}_i$ so that it has no effect over, or dependency on, the particle headings.
In all our models, this force will be
\begin{eqnarray}\label{eq:repulsion}
    \vec{F}_i &=& \mu\sum_{j \in S_i} \; 
		\frac{ \left\| \vec{r}_{ij} \right\| - R}{R} \frac{\vec{r}_{ij}}{\left\|\vec{r}_{ij}\right\|},
\end{eqnarray}
where $\vec{r}_{ij} = \vec{r}_j - \vec{r}_i$ and $\mu$ controls the force intensity.
Note that, with this definition, the alignment interaction range coincides with the repulsion range, so the repulsion and alignment forces are always simultaneously present for 
$\vec{r}_{ij} < R$.
Since repulsion vanishes linearly at $\vec{r}_{ij}=R$, however, it is a relatively weak interaction that still allows the alignment forces to dominate the collective dynamics. 
Its main role will be to favor the homogeneous states that will be compared to analytical results in the following sections.
When repulsive forces are present, we made the choice of setting the repulsion range and the alignment range equal to $R$, in order to reduce the total number of parameters and to avoid the complications of additional structures that can emerge when the interactions have an alignment-only region. 
In addition, we will also consider the case without repulsive forces, with $\vec{F}_i=0$, which is equivalent to the limit case with a vanishing interaction range for the repulsion.

\section{Analysis Tools}
\label{sec:tools}

In this section, we will introduce the nondimensional control parameters and order parameters that we will use to characterize our simulation results.

\subsection{Control parameters}

Following the approach in \cite{martin2018collective}, we compute the phase diagram as a function of two dimensionless control parameters: the Peclet number $\text{Pe}$ and the interaction strength $\text{g}$.

The Peclet number is defined in terms of the model parameters as
\begin{equation}
    \label{eq:Pe}
    \text{Pe} = \frac{v_0}{R \sigma_\theta^2}.
\end{equation} 
This represents the ratio of the advection rate over the diffusion rate.
Smaller $\text{Pe}$ values thus correspond to a more diffusive motion; larger $\text{Pe}$ values indicate that self-propulsion plays a more prominent role. 
The persistence length ($v_0/\sigma_{\theta}^2= \text{Pe} \, R$) defines the typical scale over which a noninteracting particle will lose the information of its initial orientation. 
Note that, the bigger the persistence length, the larger the simulation box required to avoid finite-size effects.

The dimensionless interaction strength is defined by 
\begin{equation}
    \label{eq:g}
   \mathrm{g}=\frac{1}{\tau \sigma_\theta^2}.
\end{equation}
It represents the ratio of the alignment rate over the angular diffusion rate and is expressed in terms of the noise level and the typical relaxation time $\tau$ that particles would take to align in the absence of noise. 
Larger $\text{g}$ values thus represent a stronger tendency to align, when considering the combined effect of the aligning forces and noise. 

\subsection{Order parameters}

We will use two order parameters to characterize the collective stationary states in the phase diagrams: the polarization $\Phi$ and the clustering $\Gamma$.

The degree of global order is described by the polarization, an orientational order parameter defined by
\begin{equation}
    \label{eq:polarization}
    \Phi=\frac{1}{N v_0} \left|\sum_{i = 1}^{N}\vec{v}_i\right|,
\end{equation}
where $N$ is the total number of particles. 
With this definition, $\Phi=1$ if all particles move in exactly the same direction and $\Phi\approx 0$ if they are heading in fully random directions.

The degree of homogeneity in the spatial distribution of particles is quantified by a clustering order parameter, which we define as
\begin{equation}
    \label{eq:clustering-factor}
    \Gamma = \frac{\sqrt{\sum N_k^2}}{N}.
\end{equation}
Here, the sum is performed over all clusters and $N_k$ is the number of particles within cluster $k$.
Each cluster $k$ is composed by all particles within the interaction range (closer than $R$) of any other particle that is also a member of cluster $k$.
With this definition, any particle that is not interacting with any other is the sole member of a cluster with $N_k=1$.
The value of $\Gamma$ varies between $1/\sqrt{N}$ and $1$.
For the low density case considered here, $\Gamma=1/\sqrt{N}$ implies that no particles are interacting and that the spatial distribution is highly homogeneous.
The $\Gamma=1$ case implies instead that all particles belong to a single cluster and that the system presents strong local density fluctuations.

%
\begin{figure*}[t]
    \begin{tabular}[b]{c}
    \begin{subfigure}[b]{1.3\columnwidth}
      \includegraphics[width=1.0\textwidth]{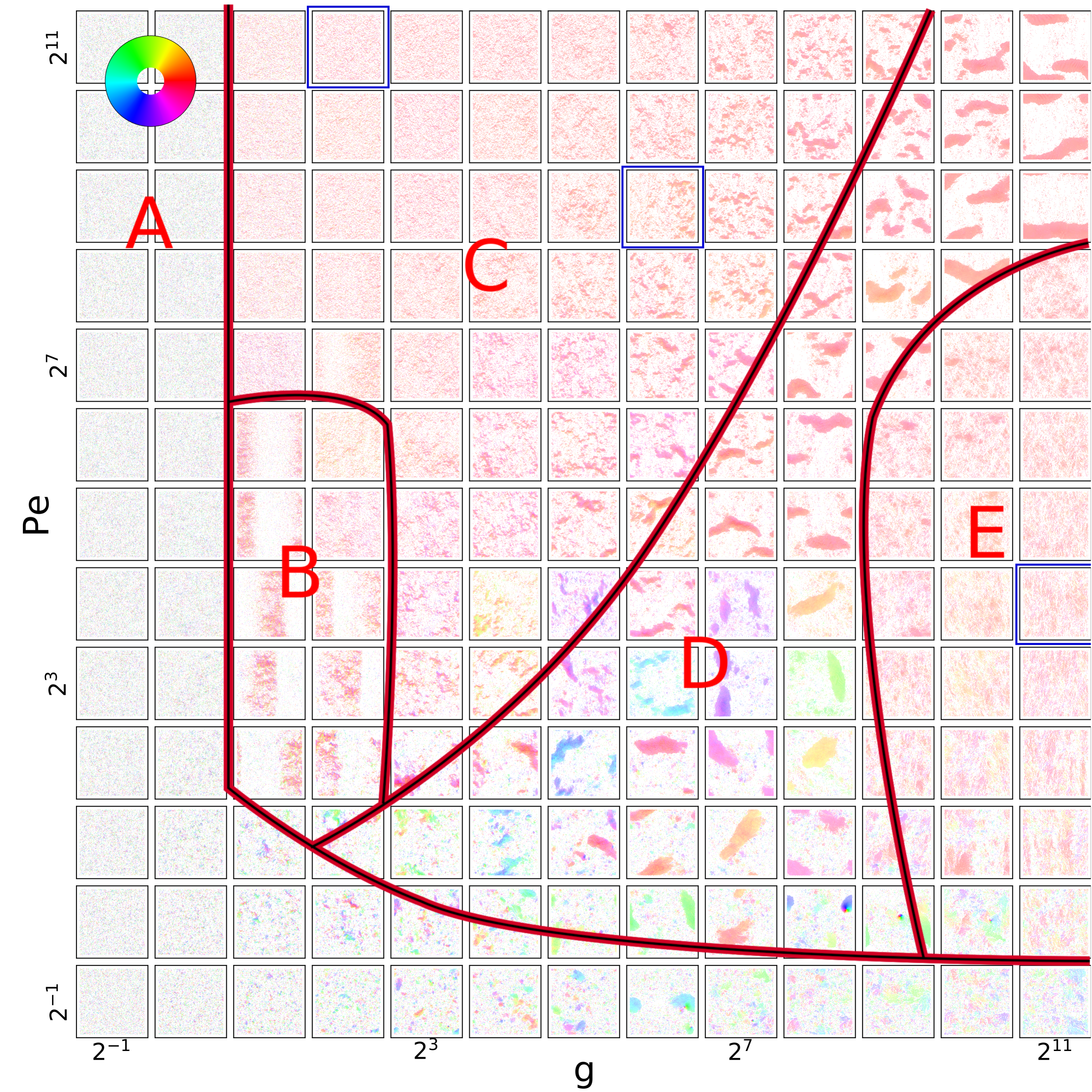}
      \caption{Snapshots phase diagram}
      \label{fig:snapshot-diagram}
    \end{subfigure}
    \end{tabular}
    \\
    \begin{tabular}[b]{cc}
      \begin{subfigure}[b]{0.8\columnwidth}
        \includegraphics[width=0.8\textwidth]{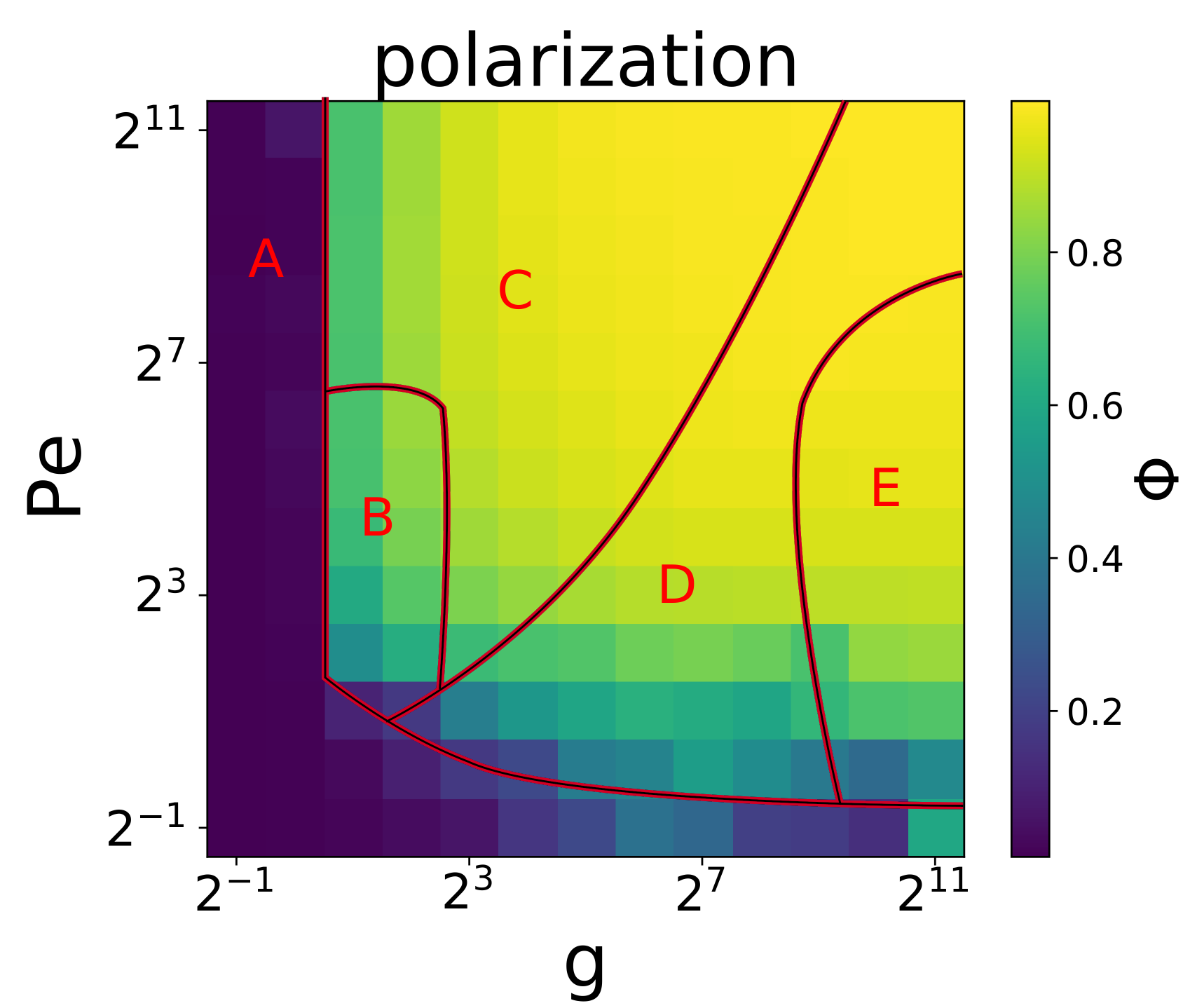}
        \caption{Polarization phase diagram}
        \label{fig:polarization-diagram}
      \end{subfigure}
      \begin{subfigure}[b]{0.8\columnwidth}
        \includegraphics[width=0.8\textwidth]{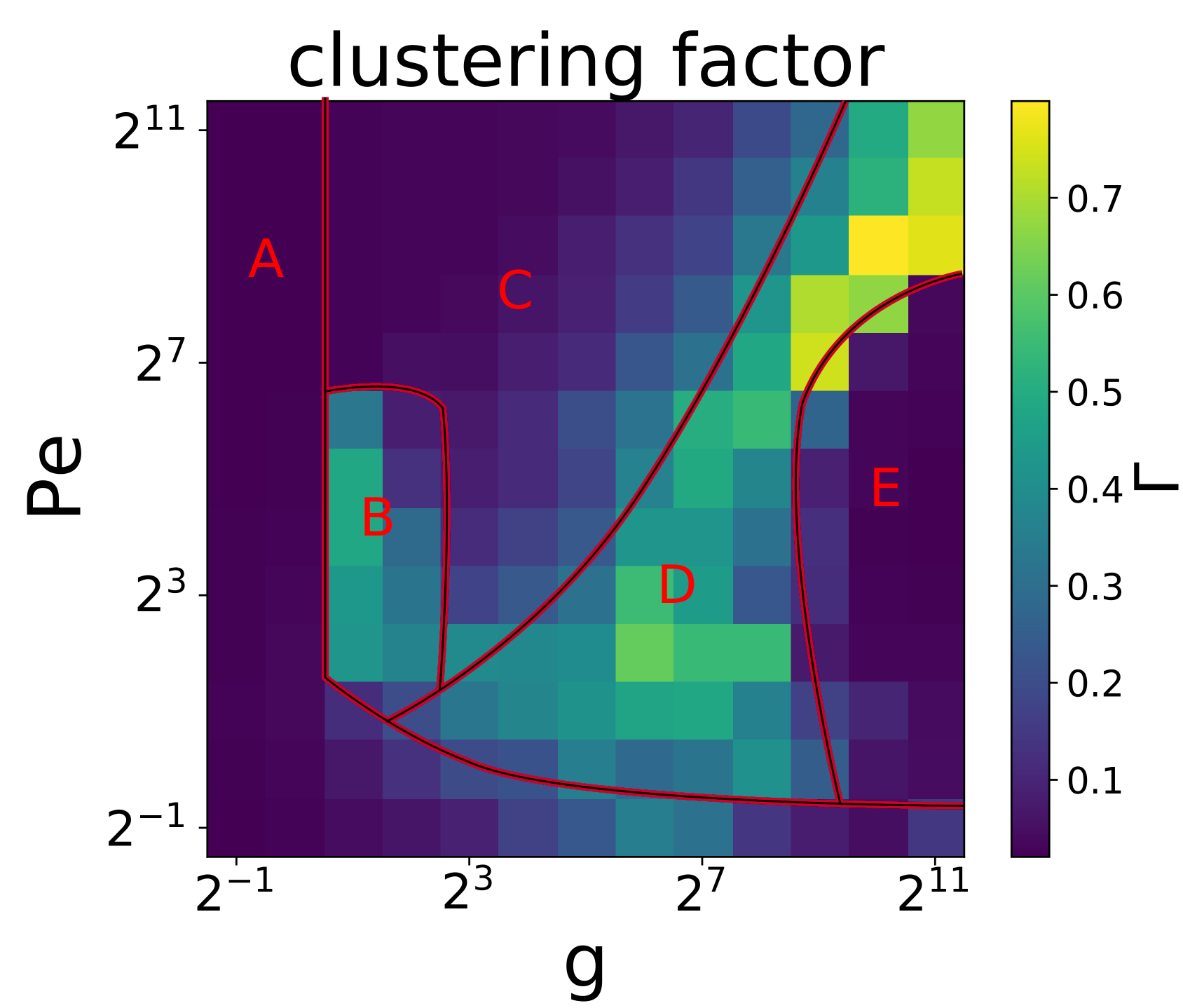}
        \caption{Clustering factor phase diagram}
        \label{fig:clustering-diagram}
      \end{subfigure}
  \end{tabular}
  \caption{\label{fig:ABC}
  Three representations of the phase diagram of the Mean-Angle model with repulsion as a function of the dimensionless control parameters (coupling strength $\text{g}$ and Peclet number $\text{Pe}$). 
  The top panel \textbf{(a)} displays representative snapshots of the stationary states for different parameter combinations. Each particle is colored according to its heading direction (as shown in the top-left color disk). Regions with homogeneous color thus indicate high orientational order.
  The snapshots with blue frames highlight the homogeneous polarized states analyzed in Section \ref{sec:analysis}.
  The bottom panels present the values of the polarization $\Phi$ \textbf{(b)} and clustering $\Gamma$ \textbf{(c)} order parameters as a function of $\text{g}$ and $\text{Pe}$.
  The A-E labels identify the different regions discussed in the main text and the red lines sketch the boundaries between them.
  }
\end{figure*}

\section{Phase space}
\label{sec:overview}

In this section, we will describe the phase space of the stationary states reached by the MA model, with the alignment dynamics given by Eq.~(\ref{eq:model-average-angle}). 
We will first specify our simulation setup and then describe, in separate subsections, the main regions of the phase space.

We begin by defining the quantities that we kept constant for all simulations. 
Without loss of generality, we fixed the length scale unit by setting $R=1$.
Note that this makes the value of $\text{Pe}$ equal to the persistence length of non-interacting particles.
In order to reduce the parameter space and focus on the alignment dynamics, we also set the self-propulsion speed to $v_0=0.2$ and the repulsion strength to $\mu = 1$. 

The mean density was set relatively low for all runs, with total packing fraction
$\eta = N \pi (R/2)^2 / L^2 = 0.3$, where $R/2$ is the effective radius of the interacting objects and
$L$ is the size of the simulation box.
We made this choice because densities near or above the packing fraction tend to produce crystallized states that we aim to avoid, since they are incompatible with the usual active hydrodynamic states that we are interested in studying.
To keep this density fixed, the size of the simulation box will be $L=(R/2) \sqrt{N \pi / \eta}$ in all runs below.

We explored the phase space within the range $\text{Pe} \in [0.5,2048]$ and $g \in [0.5, 2048]$ by changing the alignment relaxation time $\tau$ and the angular noise level $\sigma_{\theta}$, while keeping all other parameters fixed.
In order to ensure the convergence to a stationary state, each simulation was run for at least $10^6$ steps.
To adequately resolve the temporal dynamics, we used a different simulation time-step 
$\Delta t$ for different coupling strengths, such that $\Delta t \leq \tau/5$, with a maximum of $\Delta t=0.1$ for weak coupling and a minimum of $\Delta t=10^{-4}$ for strong coupling.
For the smallest time-step values, the number of simulation steps was increased, reaching up to $10^8$ steps, which corresponds to a final time $t=10^4$. 

Figure \ref{fig:ABC} displays three representations of the phase space of the MA model, as a function of $\Phi$ and $\Gamma$ in logarithmic scales. 
We simulated a system of $N = 10^4$ particles in a periodic box of side $L \approx 161.8$, using the parameter values detailed above.
Panel \ref{fig:snapshot-diagram} presents small snapshots of the final states of runs computed for each parameter combination. The particles are colored according to their heading directions. 
When a region displays collective motion, it therefore appears with a uniform color and when it has random headings, it appears in grey.
Panel \ref{fig:polarization-diagram} shows the corresponding polarization $\Phi$ and panel \ref{fig:clustering-diagram} the clustering $\Gamma$, averaged over the final $10^3$ frames of five independent simulations, after they reach their stationary states.
We overlaid on all panels the approximate boundaries between the different domains, which were identified by changes in the degree of polarization or clustering and are clearly reflected in the snapshots. Each of the resulting domains will be discussed in detail in the subsections below.

To better understand this diagram, we note that variations of the angular noise $D_\theta$, the alignment time $\tau$, or the speed $v_0$ correspond to moving along straight lines with different orientations in this log-log representation.
More specifically, we have that:
\begin{itemize}
\item Decreasing $\tau$ will increase $\mathrm{g}$ along the horizontal direction and produce stronger alignment.
This is equivalent to decreasing $\sigma_\theta$ and $v_0/R$, keeping $\text{Pe}$ constant, which results in identical dynamics if we either vary $\sigma_\theta$ and $v_0$ while rescaling time and $\mu$, or vary $\sigma_{\theta}$ and $R$ while rescaling space, time, and the box size $L$ to keep the same mean density $\eta$.

\item Decreasing $\sigma_{\theta}$ while increasing $\tau$, to keep $\text{g}$ constant, will increase $\text{Pe}$ along the vertical direction and result in a higher persistence length of non-interacting particles.
This is equivalent to increasing $v_0/R$, which results in identical dynamics if we rescale time and $\mu$, as well as $L$ to compensate for the change in $R$ and keep $\eta$ unchanged.
\item Decreasing the noise $\sigma_{\theta}$ for a fixed $\text{Pe}/\text{g}=v_0 \tau / R$ ratio will move along diagonals with slope $1$ towards higher $\text{g}$ and $\text{Pe}$ values. 
Different diagonals can be reached by changing $v_0$, $\tau$, or $R$.
\end{itemize}

We will now describe the regions identified in Fig.~\ref{fig:ABC}. 
We note that these do not correspond to standard thermodynamic phases, given that: 
(i) they are not in thermal equilibrium, although they have reached statistically stationary dynamics; 
(ii) they may be the result of finite-size effects, and could disappear in the thermodynamic limit; and 
(iii) they are not necessarily separated by phase transitions at critical boundaries, and may instead develop progressively.
Despite this, we refer here to the resulting regions as phases, in an analogy to standard thermodynamics, and will present below a qualitative overview of each one for the different models.

In the following subsections, we begin by combining the information of all panels in Fig.~\ref{fig:ABC} to characterize five different regions in the phase diagram of model MA.

\subsection{Region A: Disorder}
Region A displays no global orientational order.
It presents an ``L'' shape in the $\text{g}$-$\text{Pe}$ plane, since it appears at low $\text{g}$ or low $\text{Pe}$ values.
Two subregions can be distinguished within this region.
For low $\text{g}$, we find a homogeneous subregion with no local clustering, which corresponds to the disordered state most commonly studied in standard flocking models such as the Vicsek model.
For low $\text{Pe}$ and high $\text{g}$ values, we observe instead the formation of locally ordered clusters that move in different directions and display complex fission-fusion dynamics.
Although these clusters are typically too small to produce high polarization, we observe that, for some periods of time, large clusters can form and significantly increase the polarization, which results in the observed intermediate mean $\Phi$ values.
This may be a finite-size effect, however, since large clusters that significantly affect global order are increasingly rare in larger systems.
We thus consider here these states as part of region A, as they exhibit no robust, stationary order.

\subsection{Region B: Large-scale bands}

Region B exhibits global orientational order and an inhomogeneous density distribution with large-scale travelling bands. These bands are aligned perpendicular to the heading direction and span here the simulation arena, across the periodic boundary condition. 
Note that these band states appears only in the vicinity of the order-disorder transition.
They can be linked to a generic instability of the homogeneous ordered regime that has been identified in previous studies \cite{bertin2009hydrodynamic,ihle2013invasion,peshkov2014boltzmann}. 

\subsection{Region C: Low coupling homogeneous order}

Region C is characterized by ordered stationary states with homogeneous density. 
By analyzing simulations of different sizes, we found that this region can be divided into at least two subregions. 
The first subregion, next to region A and immediately above region B, appears homogeneous and displays no bands only because of finite-size effects. Indeed, for large $\text{Pe}$ values the persistence lengths become too big for the simulation box to contain the band structure. 
We verified that bands reemerge in the high Peclet number regime if we simulate larger systems.
In the thermodynamic limit, we thus expect region B to appear between regions A and C, even for high $\text{Pe}$ values, which is in agreement with the instability of the homogeneous ordered state close to the transition predicted in \cite{bertin2009hydrodynamic,ihle2013invasion,peshkov2014boltzmann}.
The second subregion can be found for larger $\text{g}$ values (here $g \gtrsim 2^3$) and contains relatively homogeneous ordered states that we expect will not present significant changes in the thermodynamic limit. 
It corresponds to the regime that has been analyzed theoretically using hydrodynamic and kinetic theory \cite{ToTuPRE98,toner2005hydrodynamics,ihle2011kinetic,toner2012reanalysis} and extensively studied in numerical simulations of the Vicsek model \cite{chate2008collective,mahault2019quantitative}.
We note that the degree of homogeneity changes within the region. 
For larger $\text{g}$, we observe stronger density fluctuations and the emergence of clusters of all sizes.

\subsection{Region D: Clustering}

Region D is characterized by the presence of big clusters that span the size of the system and  is identified in panel \ref{fig:clustering-diagram} by its large $\Gamma$ value.
The transition from region C to region D appears to be smooth, since clusters start forming in region C as we approach the boundary, but the clusters in region D have distinct characteristics. 
Indeed, as shown in the Supplementary Material, 
in region D the size of large clusters along the heading direction scales faster than their size perpendicular to it, and faster than the system size $L$ (for fixed mean density). 
This implies that a few elongated giant clusters will tend to dominate the dynamics in very large systems and that the length of the largest cluster along the heading direction will often exceed $L$, so it will connect to itself across the periodic boundary.
We point out that an equivalent clustering regime was discussed in detail in \cite{martin2018collective} and that the transition from a homogeneous state to this regime can be described theoretically by analyzing the stationary distribution of cluster sizes \cite{peruani2010cluster,martin2018collective}.

\subsection{Region E: Strong coupling homogeneous order}

Region E contains homogeneous, ordered states that are reached as the alignment strength $\text{g}$ is increased beyond the clustering state. In this regime, large clusters disappear because particles align almost immediately when they come into contact. This limits their overlap and, when combined with the repulsive interactions, disfavors the formation of the high density regions required to sustain large persistent clusters. 
We observe instead small clusters that, in contrast to region D, do not scale with system size and tend to be shorter along the heading direction than perpendicular to it.
We note that regions C and E thus correspond to two different types of homogeneous ordered regimes, separated by the clustering region D.

\section{Statistical properties of homogeneous polarized states}
\label{sec:analysis}

In this section, we will study in detail the homogeneous polarized states that we find within the phase diagram presented above. 
This type of states should satisfy the assumptions of the Toner-Tu theory and have thus been the focus of multiple theoretical analyses \cite{ToTuPRE98,tu1998sound,toner2005hydrodynamics,toner2012reanalysis}.
We are interested here in distinguishing which of them, if any, best matches the corresponding analytical results.

We selected two states in region C and one in region E that display high polarization $\Phi \approx 1$ and low clustering $\Gamma \approx 0$.
These are marked by blue frames in Fig.~\ref{fig:snapshot-diagram} and defined as:
state C-1 with $\text{Pe}=2048$, $g=4$ ($\Phi=0.86$, $\Gamma=0.003$);
state C-2 with $\text{Pe}=512$, $g=64$ ($\Phi=0.98$, $\Gamma=0.015$); and 
state E with $\text{Pe}=16$, $g=2048$ ($\Phi=0.94$, $\Gamma=0.003$).
Note that we distinguished two points in region C because, as discussed in Section \ref{sec:overview}, state C-1 is in a subregion that is homogeneous only due to finite-size effects whereas states C-2 and E are expected to remain homogeneous for any system size.
We will compare below the statistical properties of these three states, computing their correlation functions, giant number fluctuation, and local density-order relationship.

We begin by examining the equal-time correlation function of the velocity fluctuations, defined in a continuous field by
\begin{equation}
    \label{eq:spatial-correlation-function}
    C(\vec{r})=\left \langle \delta \vec{v}(\vec{r}_0,t) \cdot \delta\vec{v}(\vec{r}_0+\vec{r},t)\right \rangle_{\vec{r}_0,t},
\end{equation}
where the mean $\langle \cdot \rangle_{\vec{r}_0,t}$ is taken over space $\vec{r}_0$ and time $t$, after a stationary state is reached.
Here, $\delta\vec{v}(\vec{r},t)=\vec{v}(\vec{r},t)-\vec{V}(t)$ denotes the velocity fluctuations, with $\vec{v}(\vec{r},t)$ the velocity field 
and $\vec{V}(t)=\left\langle \vec{v}(\vec{r},t)\right\rangle_{\vec{r}}$ the mean velocity, averaged over space at time $t$.
We decompose this function into 
$C(\vec{r}) = C_{\parallel}(\vec{r}) + C_{\perp}(\vec{r})$ by defining
\begin{eqnarray}
    C_{\parallel}(\vec{r}) &=& 
    \left\langle 
    \delta v_{\parallel}(\vec{r}_0,t) \cdot \delta v_{\parallel}(\vec{r}_0+\vec{r},t) 
    \right\rangle_{\vec{r}_0,t}
    \\
    C_{\perp}(\vec{r})  &=& 
    \left\langle 
    \delta v_{\perp}(\vec{r}_0,t) \cdot \delta v_{\perp}(\vec{r}_0+\vec{r},t) 
    \right\rangle_{\vec{r}_0,t}
    \label{eq:corrCperpR}
\end{eqnarray}
as the correlation functions of the velocity fluctuations parallel and perpendicular to the mean heading direction, $\delta v_{\parallel}$ and $\delta v_{\perp}$, respectively.
We will focus below on $C_{\perp}(\vec{r})$, since the first order approximation of $\delta{v}$ only involves $\delta v_{\perp}$ in highly polarized states.
Finally, we will present our results in Fourier space, to reduce the noise of the resulting curves, by defining
\begin{equation}
    \label{eq:fourier-spatial-correlation-function}
    \tilde{C}_{\perp}(\vec{q}) = \left\langle 
    \tilde{\delta v}_{\perp}(\vec{q}_0,t) \cdot 
    \tilde{\delta v}_{\perp}(\vec{q}_0+\vec{q},t)
    \right\rangle_{\vec{q}_0,t}.
\end{equation}
Here, $\vec{q}_0$ and $\vec{q}=\left(q_{\parallel},q_{\perp}\right)$ are the spatial frequency vectors, $\tilde{\delta v}_{\perp}(\vec{q},t)$ is the Fourier transform of $\delta v_{\perp}(\vec{r},t)$, and the mean is computed over $\vec{q}_0$ and $t$. 
We note that the self-propulsion and local advection fields always coincide in the Toner-Tu hydrodynamic theory, but could have in principle distinct values, given by $v_0 \hat{n}_i (\theta_i)$ and $\dot{\vec{r}}_i$, respectively. However, here due to symmetry of the repulsion force and the corresponding displacements, they are equivalent on the coarse-grained level.
We used the former in the results presented below but have also verified that we observe no significant differences when using the latter.
The numerical details of the evaluation of $\tilde{C}_{\perp}(\vec{q})$ are provided in the Supplementary Material.
In order to gather better statistics, we simulated larger systems of $1.28 \times 10^6$ particles for this analysis, while keeping the same mean density 
$\bar{\rho} = N\pi^2/(4L^2) = 0.3$.

\begin{figure}[t]
    \includegraphics[width=\columnwidth]{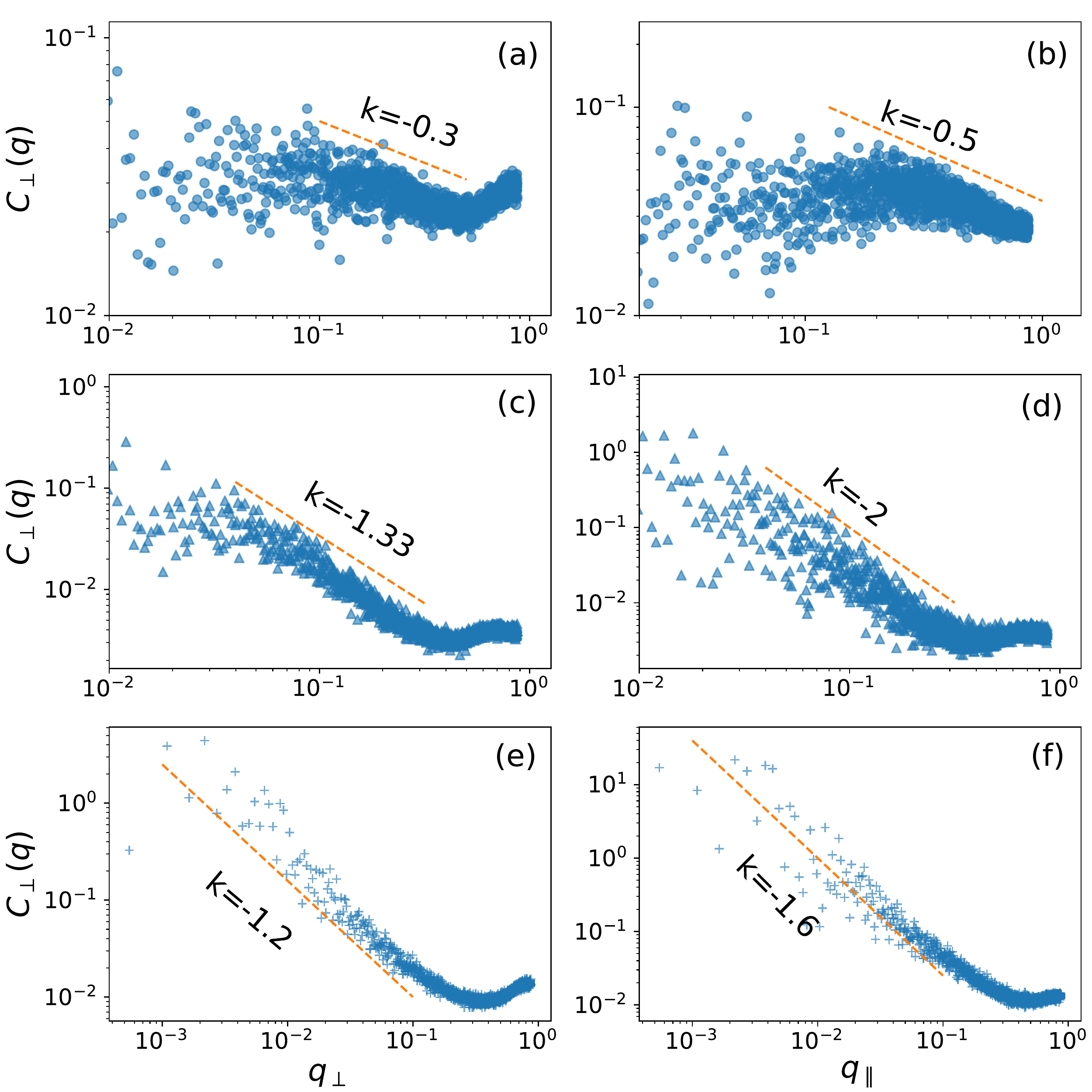}
    \caption{Correlation functions in Fourier space of the perpendicular velocity fluctuations (with respect to the mean heading direction) as a function of the perpendicular and parallel frequency vectors (left and right columns, respectively) for the three homogeneous polarized states analyzed in Section \ref{sec:analysis}:
    state C-1 (a,b), state C-2 (c,d), and state C-3 (e,f).
    The orange dashed lines with different $k$ exponents have been added for reference.}
    \label{fig:correlation-function}
\end{figure}

Figure \ref{fig:correlation-function} displays the resulting $\tilde{C}_{\perp}(q_{\perp})=\tilde{C}_{\perp}(0, q_{\perp})$ and $\tilde{C}_{\perp}(q_{\parallel}) = \tilde{C}_{\perp}(q_{\parallel},0)$ 
correlation functions for states C-1, C-2, and E in the MA model.
The Toner-Tu theory predicts that, regardless of model details,  $\tilde{C}_{\perp}(q_{\perp})$ and $\tilde{C}_{\perp}(q_{\parallel})$ should follow power-laws with exponents 
$k = -1.2$ and $k = -2$, respectively
\cite{tu1998sound,toner2005hydrodynamics,toner2012reanalysis}.
Despite presenting significant fluctuations, the figure clearly shows whether the numerical curves are compatible or not with these exponents.

Panels (a) and (b) show that state C-1 strongly deviates from the theoretical prediction, displaying curves that are compatible with much more shallow exponents.
It thus appears to correspond to a very different type of homogeneous ordered state than that analyzed in the Toner-Tu theory. 
This is consistent with our observation in Section \ref{sec:overview} noting that the homogeneity of state C-1 is the result of finite-size effects, since the shallow exponents may reflect the presence of correlations that span the scale of the simulation arena.

Panels (c) through (f) show that states C-2 and E partially match the Toner-Tu predictions in different ways. 
In state C-2, we have $k_{\parallel} \approx -2$, which coincides with the theory, and $k_{\perp} \approx -1.33$, which is slightly steeper than predicted but matches the numerical results in \cite{mahault2019quantitative}.
In state E, we have $k_{\parallel} \approx -1.6$, which is shallower than predicted, and $k_{\perp} \approx -1.2$, which agrees with the prediction.
We hypothesize that the cause of these agreements and differences is found in the underlying clustering structures.
Indeed, since clusters are elongated along the mean velocity in state C-2 and perpendicular to it in state E, these directions will provide longer homogeneous regions, which is consistent with the fact that these are the same directions that best match the predicted exponents for each case.

\begin{figure}[t]
    \includegraphics[width=0.95\columnwidth]{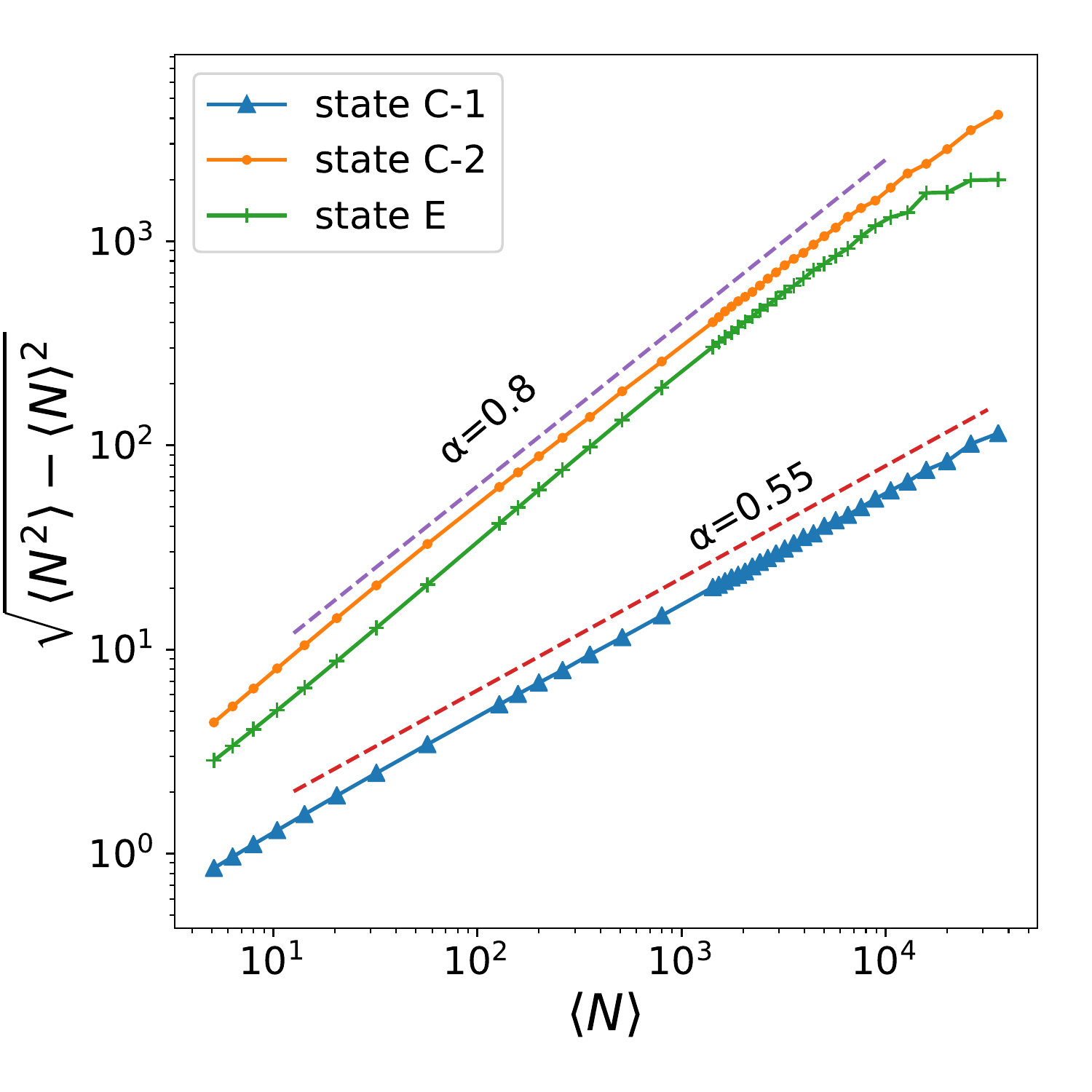}
    \caption{Standard deviation as a function of the mean number of particles in boxes of growing size for the three homogeneous polarized states analyzed in Section \ref{sec:analysis}.
    Dashed lines with different $\alpha$ exponents are added for reference.
    States C-2 and E display clear giant number fluctuations that match the theory, whereas state C-1 does not, since its $\alpha$ exponent is only slightly above $1/2$.}
    \label{fig:giant-number-fluctuation}
\end{figure}

We now turn our attention to the giant number fluctuations. These are defined as variations of the number density with a standard deviation that scales as
\begin{equation}
    \label{eq:giant-number-fluctuation}
    \sqrt{\langle N^2\rangle - \langle N \rangle ^2} \propto 
    \langle N \rangle ^ {\alpha}
\end{equation}
and with $\alpha>1/2$. Giant number fluctuations are a common feature in systems of self-propelled particles \cite{mahault2019quantitative,giavazzi2017giant,chate2008collective,ginelli2016physics}. The Toner-Tu theory predicts their presence in the homogeneous ordered state, with $\alpha=0.8$.
In order to test if this exponent holds for the three homogeneous polarized states selected, we measured the standard deviation of the number of particles in boxes of growing size in our numerical simulations. Here again, we used a larger system of $1.28\times 10^6$ particles.

Figure \ref{fig:giant-number-fluctuation} shows that all the selected homogeneous ordered states display giant number fluctuations, but with different $\alpha$ exponents.
While states C-2 and E match well the Toner-Tu prediction, state C-1 displays a much lower slope (although the slope of state C-1 can fluctuate significantly from run to run, sometimes reaching values as high as $\alpha \approx 0.65$).
These results provide further evidence that states C-2 and E can be consistent with the Toner-Tu theory but state C-1 is not. The lower exponent in the C-1 case can also be interpreted as resulting from a finite-size effect where the simulation arena may be too small to develop the processes leading to giant number fluctuations.

\begin{figure}[t]
    \includegraphics[width=\columnwidth]{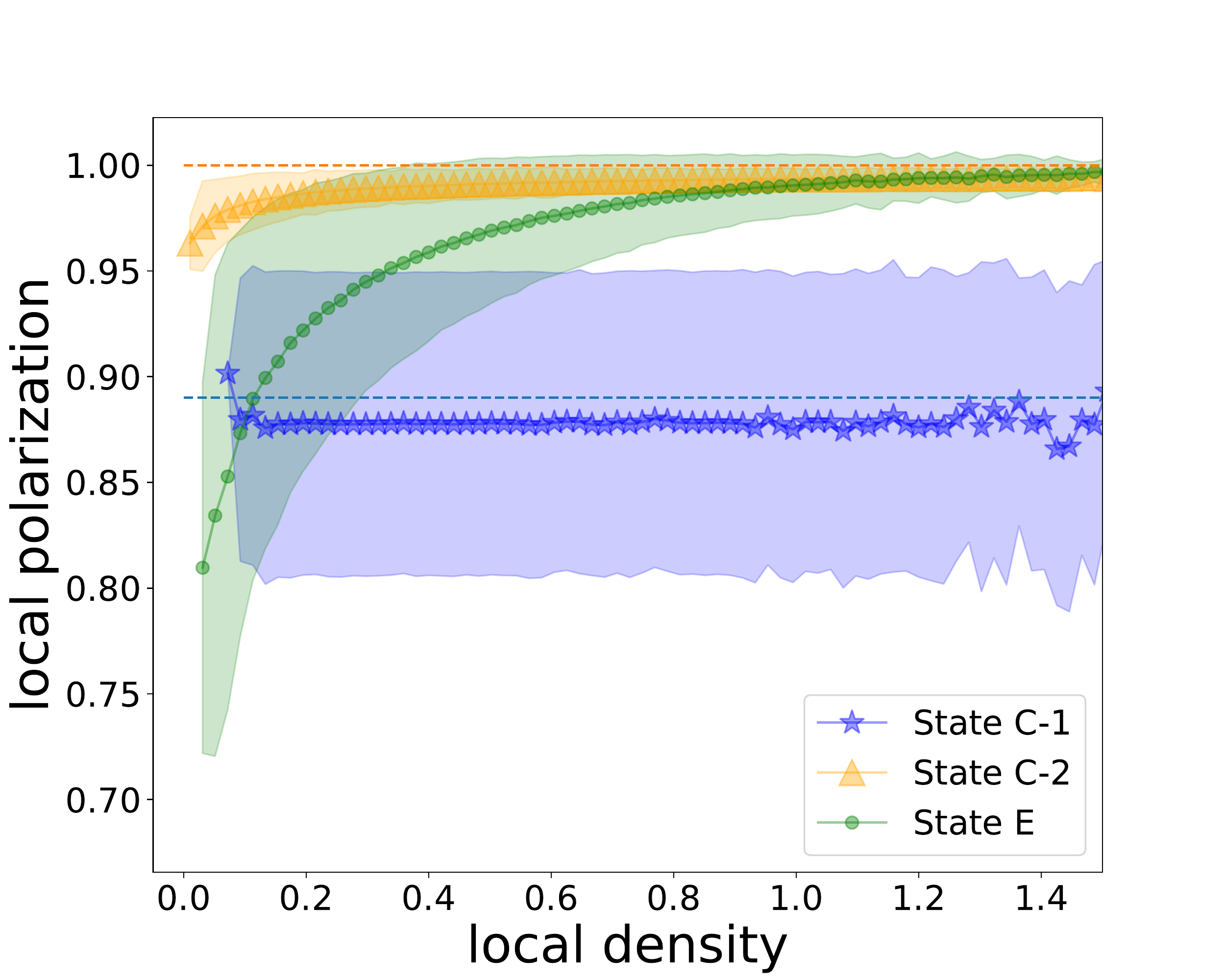}
    \caption{Local density-order correlation for the three homogeneous polarized states analyzed in Section \ref{sec:analysis}. Each curve presents the mean local polarization in regions with a given local density. 
    The lighter colored area displays its standard deviation.
    States C-2 and E show a clear local density-order correlation but state C-1 has almost constant polarization for all densities.}
    \label{fig:density-vs-polarization}
\end{figure}

\begin{figure*}[t]
    \includegraphics[width=\textwidth]{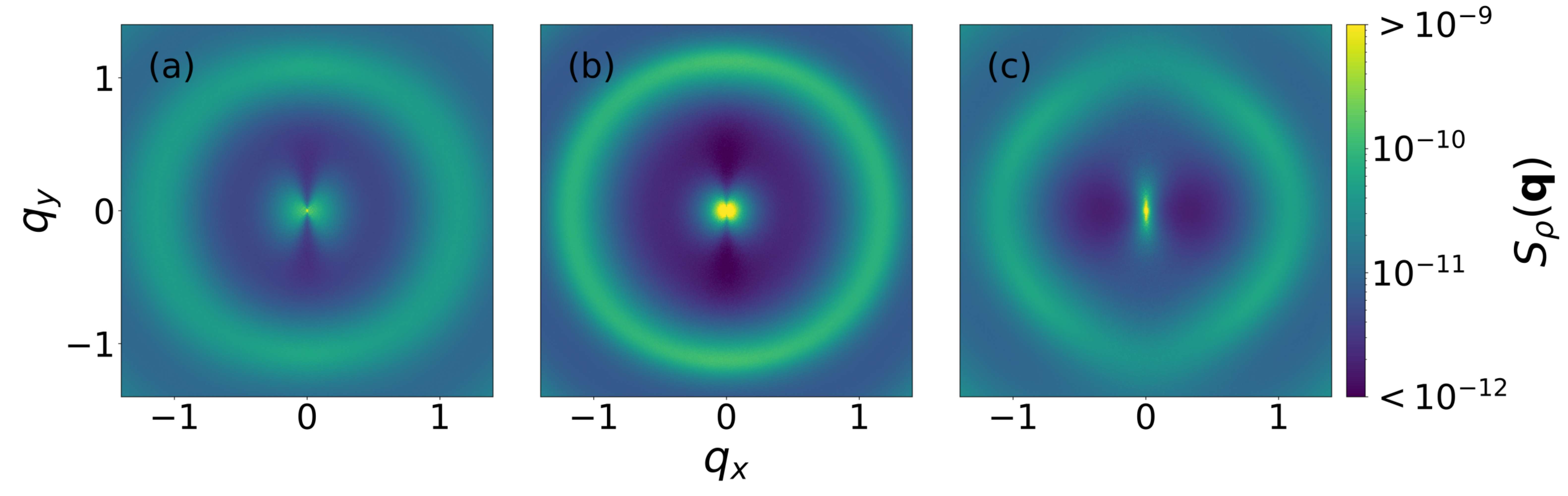}
    \caption{Fourier transform of the two-dimensional density autocorrelation function for the three homogeneous polarized states analyzed in Section \ref{sec:analysis}; states C-1 (a), C-2 (b), and E (c).
    The value of $S_{\rho}(\vec{q})$ is computed using Eq.~(\ref{eq:S(q)}).
    The mean heading direction corresponds to $+q_x$.
    All cases display a ring at $|| \vec{q} || = 1$, resulting form the $R=1$ alignment and repulsion range, but other characteristic features display significant differences.}
    \label{fig:structure-factor}
\end{figure*}

In order to further characterize the different homogeneous ordered states, we also measure the correlation between local density and local polarization.
To do this, we first find the $5$ nearest neighbors of each particle and use their positions and velocities to evaluate the local density and polarization, as detailed in the Supplementary Material. We then compute the mean local polarization and its standard deviation for each bin of values of the local density. For these calculations, we used the last $100$ frames of simulations with $3.2 \times 10^5$ particles.

Figure \ref{fig:density-vs-polarization} shows the resulting local density-order correlation for the three homogeneous ordered states considered here.
In states C-2 and E, the mean local polarization increases monotonically with the local density, that is, order is enhanced at higher densities.
This is a common situation in alignment models with metric (distance dependent) interactions, where the feedback between density and order produces persistent spatial structures in which higher density results in averaging over more particles and thus in better convergence to alignment and higher polarization.
We note that the quantitative difference between the curves for state C-2 and state E is due to their specific chosen parameter combinations and is not representative of a consistent difference between other states near C-2 and E.
State C-1 behaves here, again, very differently, since the local polarization remains constant and appears to display no correlation with the local density.
This can be understood by noticing that the particles in this state have high speed and slow alignment dynamics, which implies that the density-order feedback does not have time to develop within localized structures in the simulation arena, due to the aforementioned finite-size effects.
Each particle senses instead the mean heading of the agents within the different density regions it traverses.
The alignment dynamics thus resembles a mean-field description of the system, showing no correlation with the local density. 

Our final analysis of the homogeneous polarized states will focus on the spatial density distribution, given that the simulation snapshots show that they all display some degree of clustering.
We thus examine the two-dimensional density autocorrelation function, 
given by
\begin{eqnarray}
\label{eq:S(q)}
    S_{\rho}(\vec{q}) &=& \frac{1}{\int C_{\rho}(\vec{r}) d\vec{r}}\int C_{\rho}(\vec{r})e^{-2\pi i\vec{q}\cdot\vec{r}}d\vec{r} \\
    C_{\rho}(\vec{r}) &=& \left\langle \rho(\vec{r}_0,t) \, \rho(\vec{r}_0+\vec{r},t) \right\rangle_{\vec{r}_0,t}.
\end{eqnarray}
Here, $\rho(\vec{r},t)$ describes the local density field at position $\vec{r}$ and time $t$, obtained numerically thorough spatial binning (see Supplementary Material), and the mean $\langle \cdot \rangle_{\vec{r}_0,t}$ is taken over space and time after the stationary state is reached. For these computations we used again the last $100$ frames of simulations with $3.2 \times 10^5$ particles.

Figure \ref{fig:structure-factor} shows $S_{\rho}(\vec{q})$ for the three states considered. The heading direction corresponds here to $+q_x$.
All panels display a similar circle of radius $\vec{q}=1$, which is a consequence of the repulsion range $R=1$ determining a typical minimal distance between particles. We note that this circle is more blurry for state C-1 than for state C-2, due to the lower interaction strength $\text{g}$. 
For state E, it also appears to be less circular (Fig.~\ref{fig:structure-factor}c), which is a signature of the emergence of structures in the spatial distribution within clusters.
Near the origin, we observer that $S_{\rho}(\vec{q})$ is higher in the $\pm x$ direction than in the $\pm y$ direction for states C-1 and C-2, while the opposite is true for state E.
This reflects the presence of large-scale density structures of clusters that are typically elongated along the heading direction in states C-1 and C-2, but perpendicular to it in state E.

In sum, the results presented above show that the three homogeneous ordered states have distinctive characteristics when analyzed in detail. Importantly, none of them seems to fully match the Toner-Tu theoretical predictions, at least in the systems of over a million particle considered here. 

We found that state C-1 differs the most from the Toner-Tu theory. This can be explained by the fact that its homogeneous nature is the result of finite-size effects, as discussed in Section \ref{sec:overview}, and thus inconsistent with the infinite domain assumed by the theory.
In this state, the low $\text{g}$ and high $\text{Pe}$ values make the interactions too weak and the self-propulsion speeds too high to produce relevant density-velocity couplings at the scale of the system.
Instead, the typical persistence length is comparable to the system size and the particles appear to display quasi ballistic motion at the scale of the arena, thus resembling a gas and approaching mean-field behavior. This explains all our observations: the shallow correlation functions in the velocity fluctuations, the reduced giant number fluctuation (with $ \alpha \approx 0.5)$, and the lack of correlations between the local polarization and local density.

States C-2 and E match better the Toner-Tu theory, each in a different way, although neither reproduces all the predicted exponents. We found that both match fairly well the expected giant number fluctuations, but state C-2 displays a slightly steeper correlation function exponent $k_{\perp}$ and state E a significantly shallower $k_{\parallel}$ than predicted by the theory. 
We hypothesize that these deviations are related to the presence of anisotropic cluster structures that are still apparent in the selected states, despite being among the most homogeneous in our phase diagram.
Given the presence of giant number fluctuations at all scales, it is an open question if these structures (and, therefore, these deviations from the predicted exponents) will remain in larger systems.

\section{Comparative analyses of the phase space}
\label{sec:comparison}

We will now compare the phase diagrams and typical states obtained for variations of the MA model. 
First, we will analyze the three variations introduced in Section \ref{sec:models} (the MS, AS, and SV models), all of which satisfy similar alignment dynamics while following different specific interaction rules.
We will then describe how the phase diagrams change without repulsive interactions.

\begin{figure}[t]
    \includegraphics[width=0.23\textwidth]{figures/diagrams/AlignmentNew3/polarization-l-l.pdf}
    \includegraphics[width=0.23\textwidth]{figures/diagrams/AlignmentNew3/clustering_factors_small-l-l.pdf}
    \includegraphics[width=0.23\textwidth]{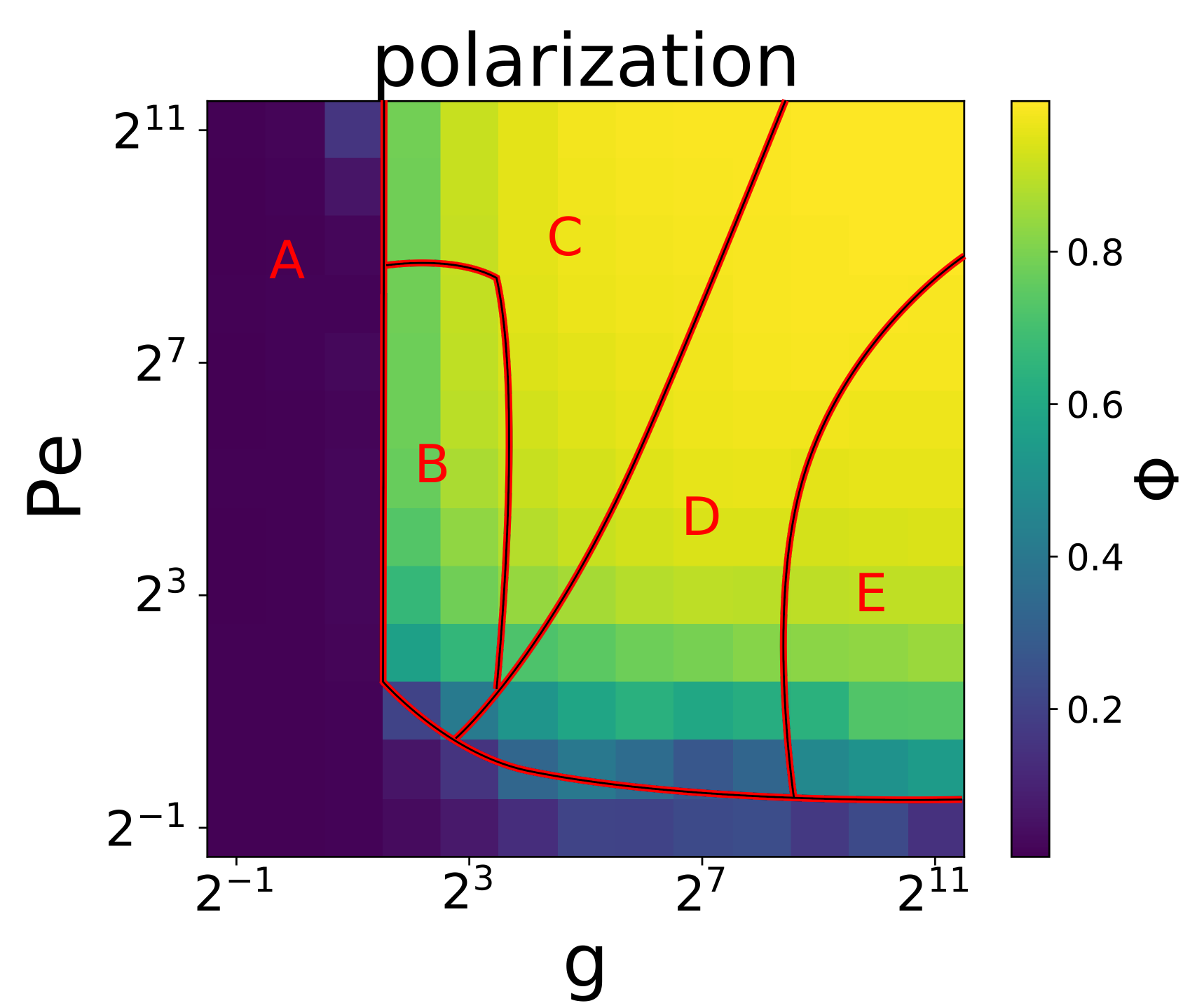}
    \includegraphics[width=0.23\textwidth]{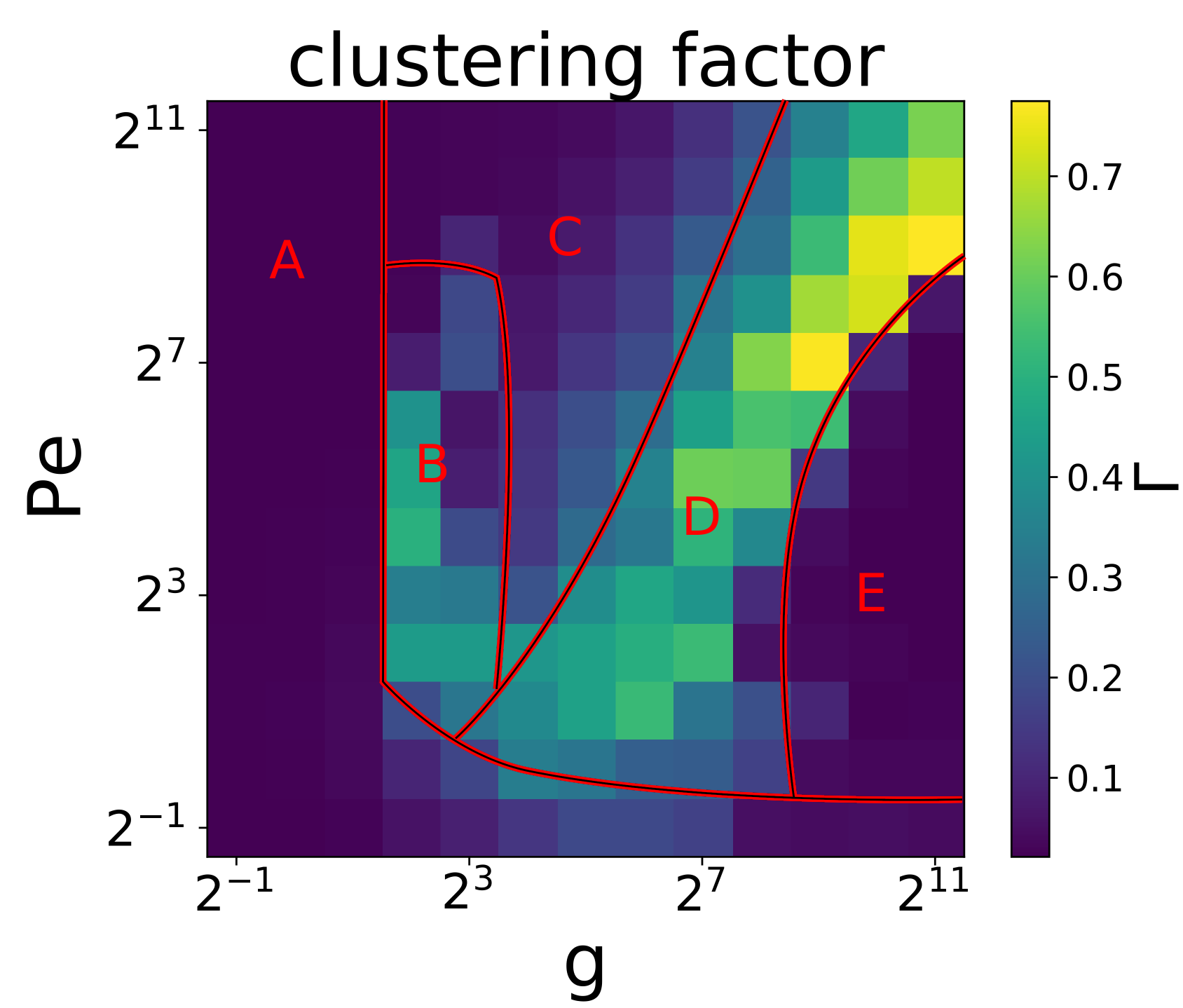}
    \includegraphics[width=0.23\textwidth]{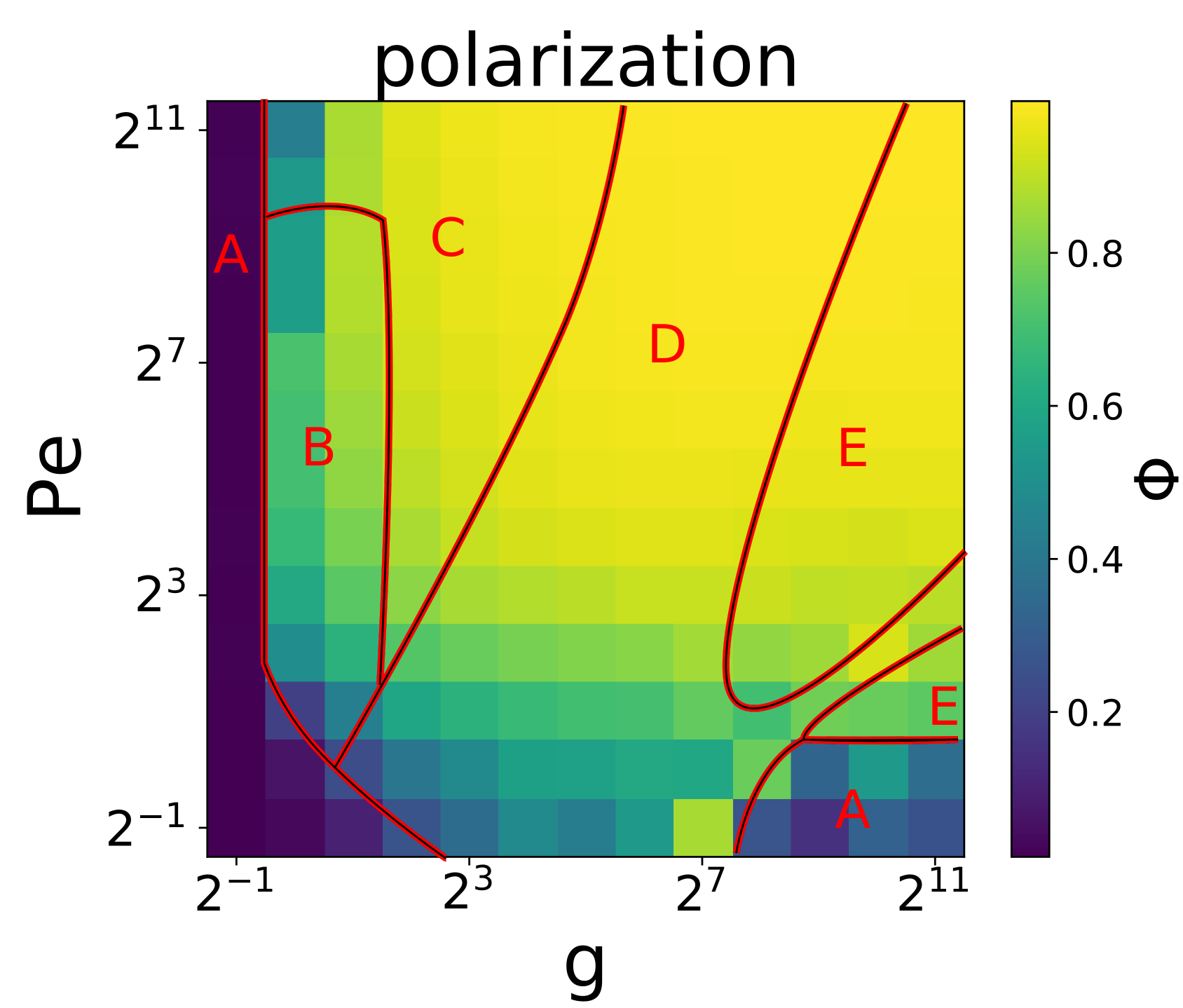}
    \includegraphics[width=0.23\textwidth]{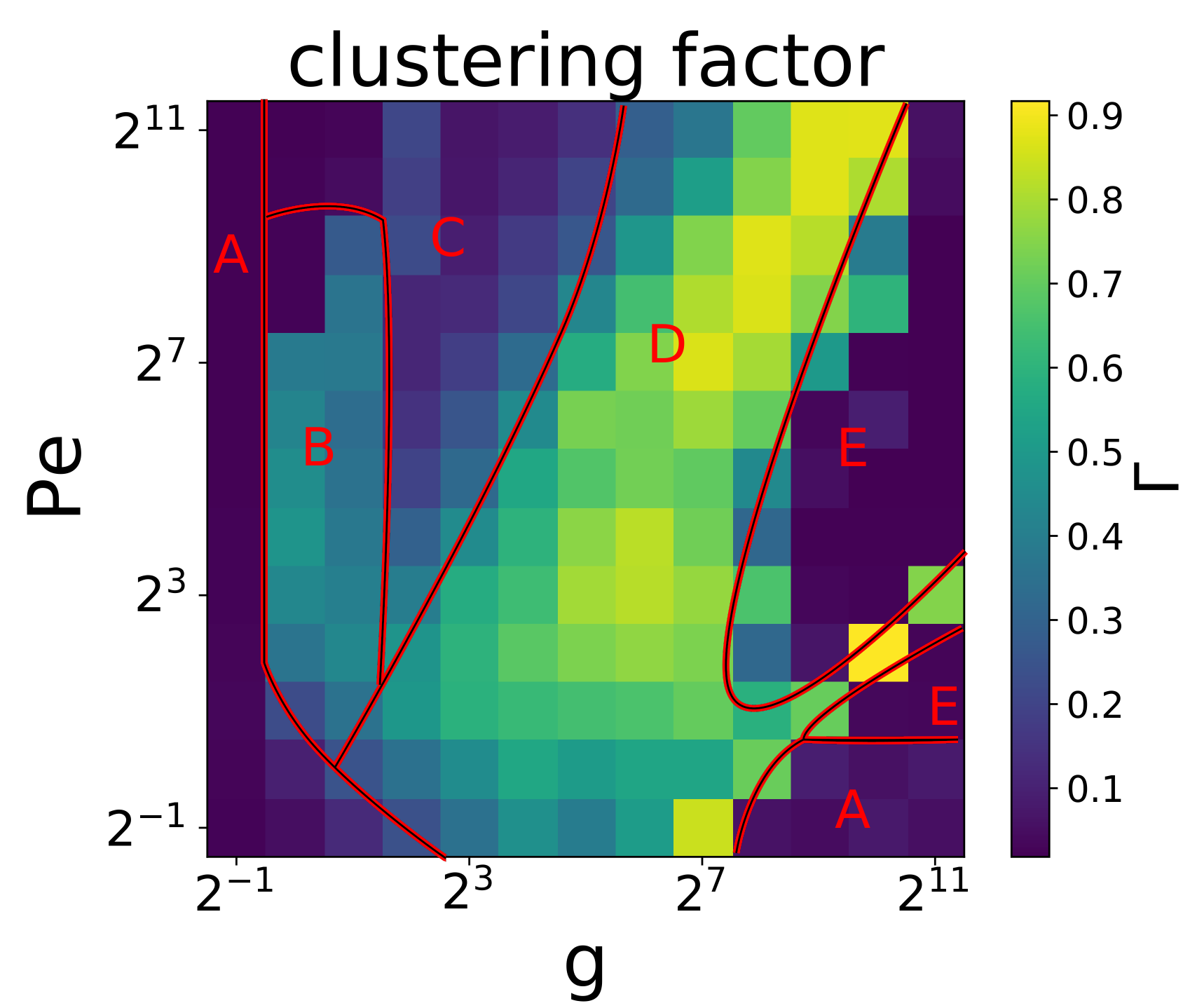}
    \includegraphics[width=0.23\textwidth]{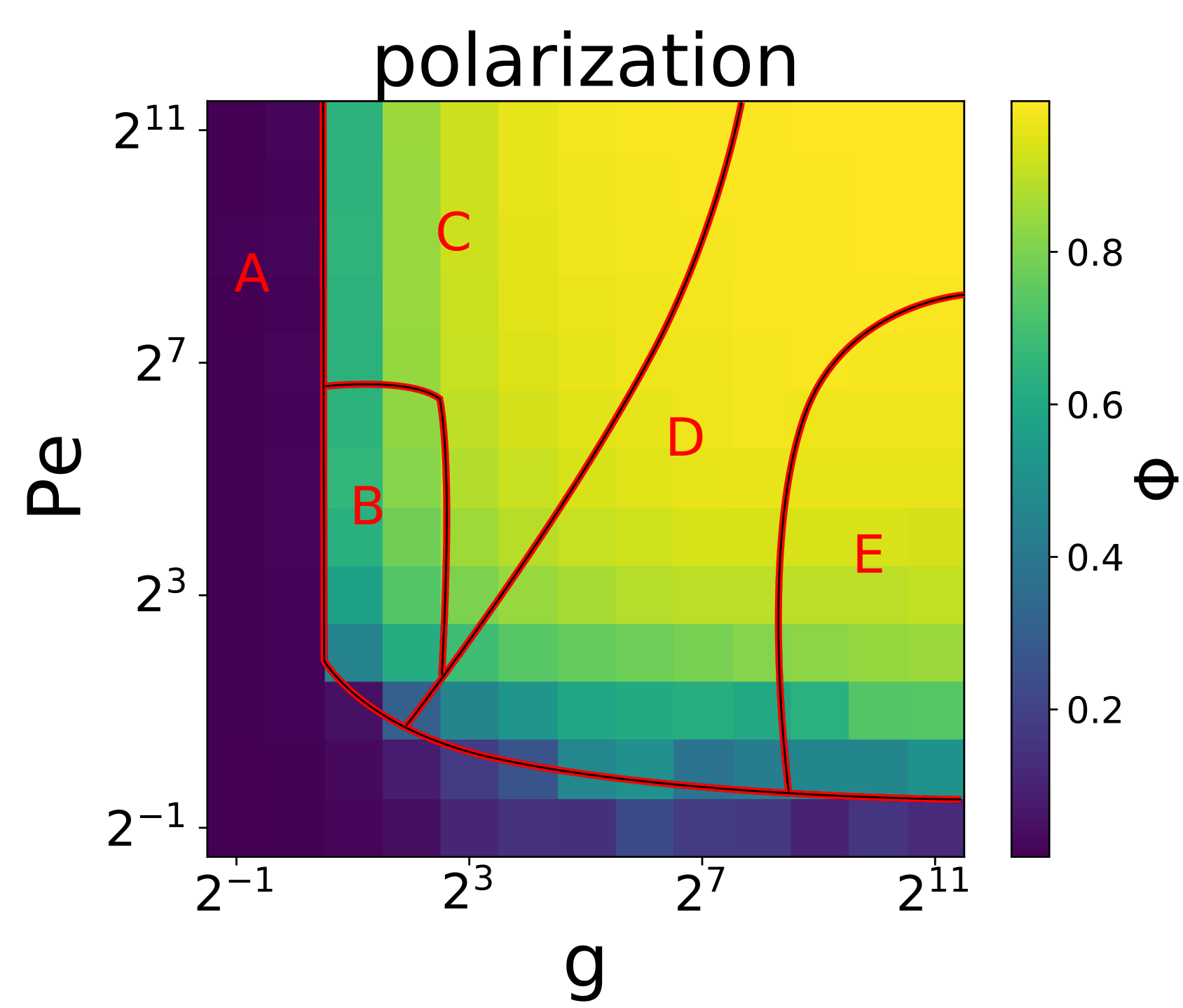}
    \includegraphics[width=0.23\textwidth]{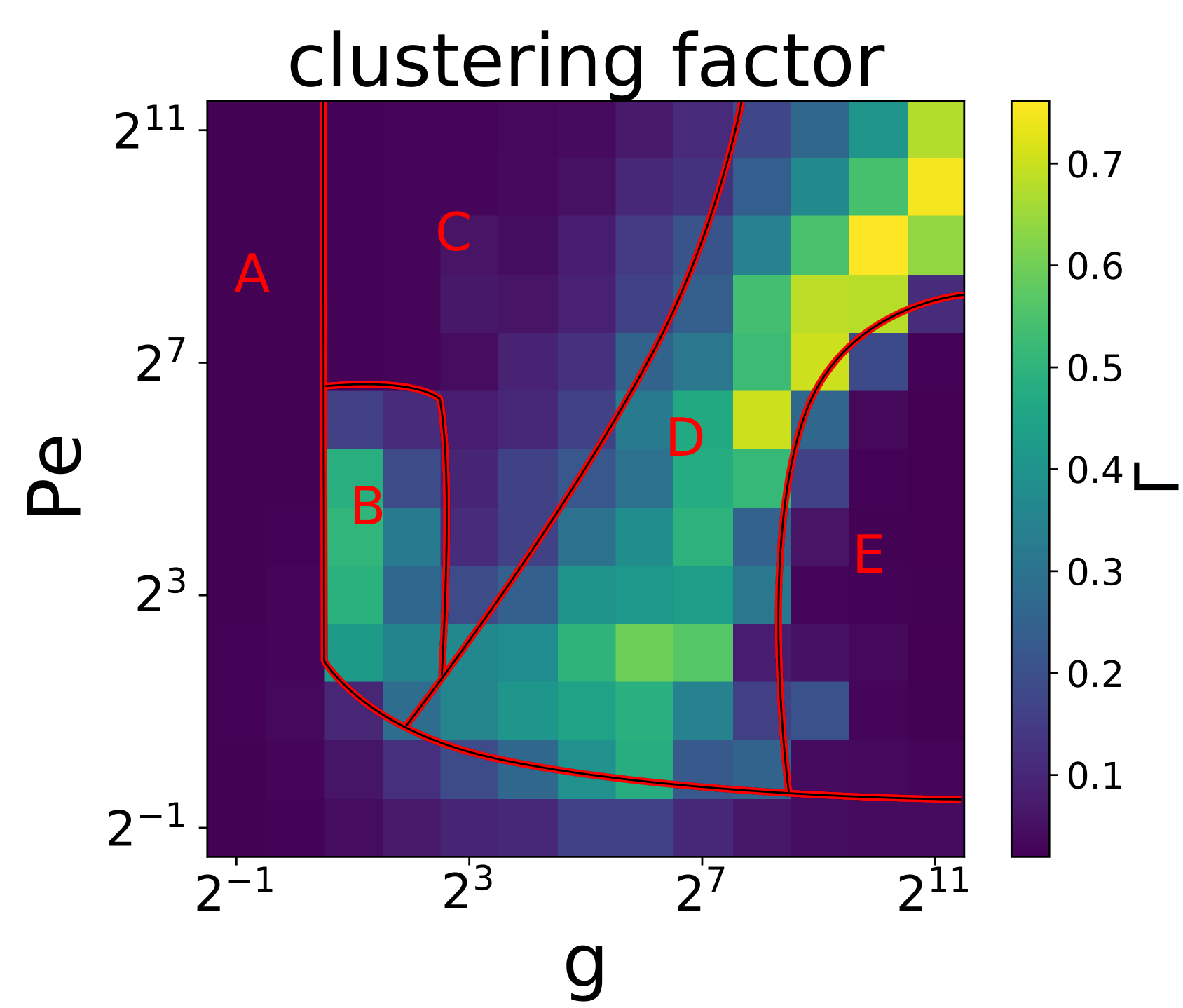}
    \caption{Phase diagrams of the polarization $\Phi$ and clustering $\Gamma$ order parameters (left and right columns, respectively) for the four continuous alignment models, with repulsive interactions, introduced in Section \ref{sec:models}. 
    Each row displays, from top to bottom, the diagrams for the Mean-Angle (same as on Fig.~\ref{fig:ABC}), Mean-Sine, Additive-Sine, and Sine-Velocity models.
    The A-E labels identify the different regions and the red lines sketch the boundaries between them.}
    \label{fig:clustering-factor&polarization-comparison}
\end{figure}

\begin{figure*}
\centering
\includegraphics[width=0.85\textwidth]{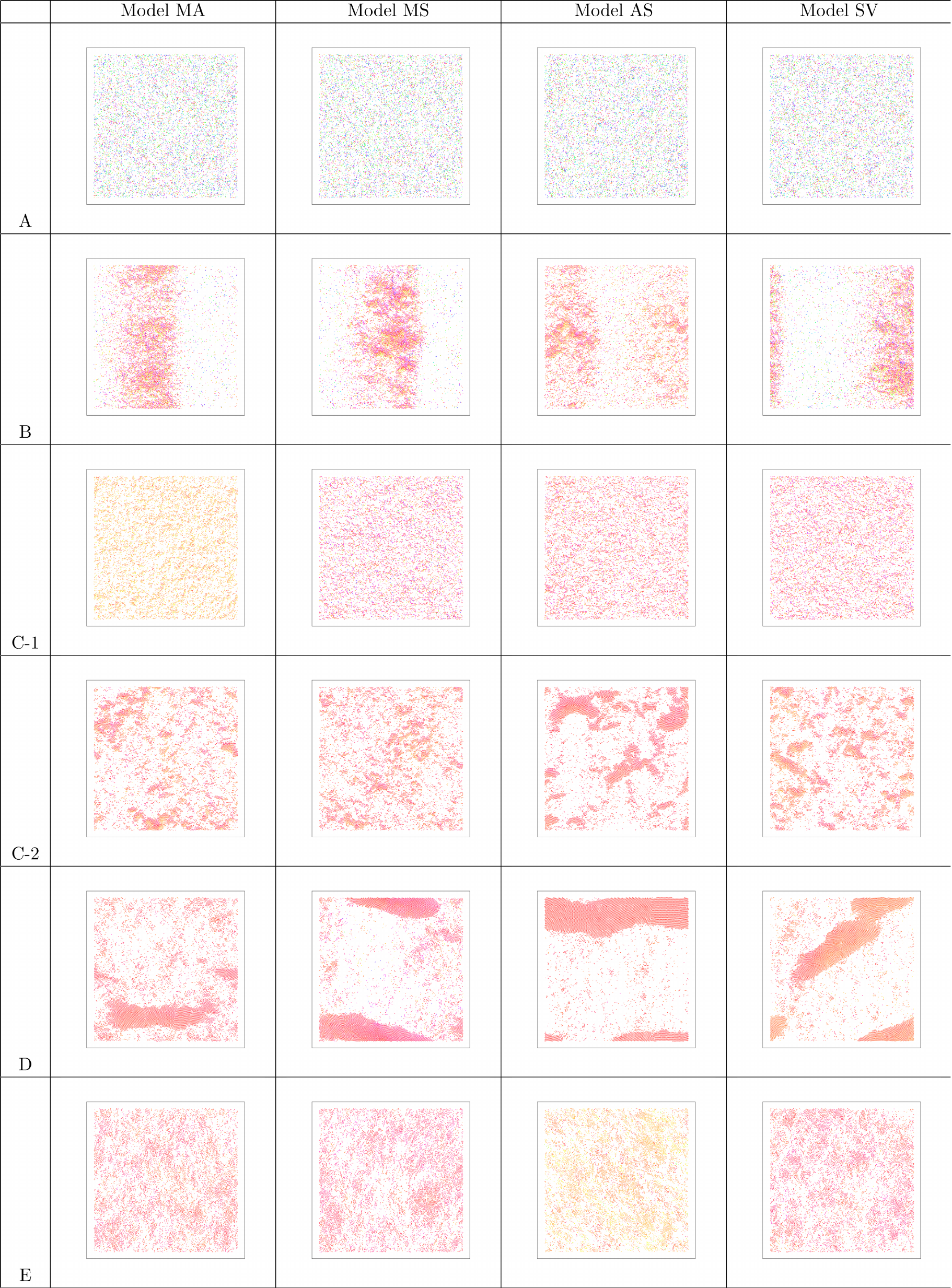}
\caption{Snapshots of representative states of the different phase diagram regions shown in Fig.~\ref{fig:clustering-factor&polarization-comparison} for the four continuous alignment models, with repulsive interactions, introduced in Section \ref{sec:models}.
Each particle is colored according to its heading direction, as in 
Fig.~\ref{fig:snapshot-diagram}.
The snapshots of equivalent regions display similar features for all four models, while differing in specific details.}
\label{fig:snap-shot-models-with-repulsion}
\end{figure*}

\subsection{Comparative description of the four models}

We present here phase diagrams for the MS, AS, and SV models detailed in Eqs.~(\ref{eq:model-average-sine}), (\ref{eq:model-sum-sine}), and (\ref{eq:model-sine-average}), respectively, comparing them to the MA case studied in previous sections.
For each model, we performed 5 independent simulations and collected the last 1000 frames, after the order parameters had reached their stationary values. As in Section \ref{sec:overview}, the following results were obtained by averaging over the $5 \times 1000$ frames selected for each model.

Figure \ref{fig:clustering-factor&polarization-comparison} presents the phase diagrams obtained for all four models in terms of the polarization and clustering factor introduced in Eqs.~(\ref{eq:polarization}) and (\ref{eq:clustering-factor}), respectively.
For comparison, the first row shows the same MA diagrams displayed in  Fig.~\ref{fig:snapshot-diagram}.
Our first observation is that their overall structure is quite similar, despite the differences in the way each algorithm tends to align the local velocities.
All the regions identified in Section \ref{sec:overview} are present in all cases, and cover similar areas of the phase diagram. 
We note, however, several differences in the details, which we describe below.

The first and second rows of the figure show that the MS and MA phase diagrams are very similar, with their main difference being that the boundaries between regions A, B, and C, are displaced.
This is consistent with the fact that the $(\theta_j - \theta_i)$ term in Eqs.~(\ref{eq:model-average-angle}) and (\ref{eq:model-average-sine}) will typically be large in the disordered region, so $\Omega_{MS}$ will be smaller than $\Omega_{MA}$ and the system will have a lower tendency to align in the MA model, which therefore displays a larger region A.
Instead, in the highly polarized regions we have
$\mathrm{mod^*}(\theta_j - \theta_i) \approx \sin(\theta_j -\theta_i)$, so the diagrams will be almost identical.

The AS diagrams in the third row of the figure are the most different from all others.
In this model, the interaction strength is not normalized by the number of neighbors, so the alignment force will be generally stronger and increase with higher local density.
This reduces the size of the disordered region A and enhances the density-alignment feedback, thus increasing inhomogeneities, which results in a larger region B and D.
In the high $\text{g}$ regime, this effect is stronger and it even changes the structure of the displayed phase diagrams by splitting regions A and E. These observations are consistent with those of \cite{chepizhko2021revisiting, kursten2021quantitative}.

Finally, the SV diagrams in the fourth row are the most similar to the MA case studied in previous sections. Here again, we expect the dynamics to match closely in the ordered regime, where the typical differences between heading angles are small. 
Interestingly, in this case we find that the low alignment regions also match well the MA model results.

Figure \ref{fig:snap-shot-models-with-repulsion} compares representative states within each region of the phase diagram for the four models. 
From top to bottom, we present typical snapshots of:
disordered states (row A), perpendicular band states (row B), finite-size homogeneous states (row C-1), low-g homogeneous states (row C-2), large longitudinal cluster states (row D), and high-g homogeneous states (row E).
Note that the snapshots selected in each region do not correspond to exactly the same $\text{Pe}$ and $\text{g}$ values for the different models, since the region boundaries change and their most representative states are displaced.
For this figure, we selected the following ($\text{Pe}$, $\text{g}$) parameter combinations. 
For the MA model A:(256,1), B:(32,2), C-1:(2048,16), C-2:(128,32), D:(256,1024), E:(64,2048);
for the MA model A:(512,1), B:(16,4), C-1:(1024,4), C-2:(256,32), D:(32,256), E:(64,2048);
for the AS model A:(512,0.5), B:(128,2), C-1:(1024,2), C-2:(2048,64), D:(1024,1024), E:(32,512);
and for the SV model A:(256,1), B:(16,2), C-1:(1024,4), C-2:(256,64), D:(128,512), E:(64,1024).

The figure shows that the general features of each characteristic state are very similar across models, although we do observe some small differences. For example, the bands in the states within region B appear to be less defined in the AS model. This may be because the enhanced density-alignment feedback discussed above favors local clusters instead of the larger band structures. We also observe that, in general, the AS model displays stronger small scale clustering in regions B, C-2, and D.
It is important to point out, however, that although some of the differences between models may appear consistently when comparing the snapshots in the figure, they may still be due to their different exact locations in the phase diagrams. Extracting general conclusions on the detailed spatial distributions favored by different models will therefore require further investigations.

\subsection{Comparing to models without repulsion}

We now study how the phase diagrams will change if there is no repulsion between particles, that is, for $\mu = 0$ in Eq.~(\ref{eq:repulsion}). 
This will connect our results to the many studies in the literature that consider self-propelled point particles without repulsion \cite{VicsekZafeirisReview2012,vicsek1995novel,gregoire2004onset,bertin2009hydrodynamic,peshkov2014boltzmann,chate2008collective,mahault2019quantitative}. 
Our simulations for this case are less exhaustive, since they only aim to identify the main differences with respect to the $\mu = 1$ models discussed above. Note that they also have increased computational cost, due to the formation of high density clusters.

\begin{figure*}
    \centering
    \includegraphics[width=0.9\textwidth]{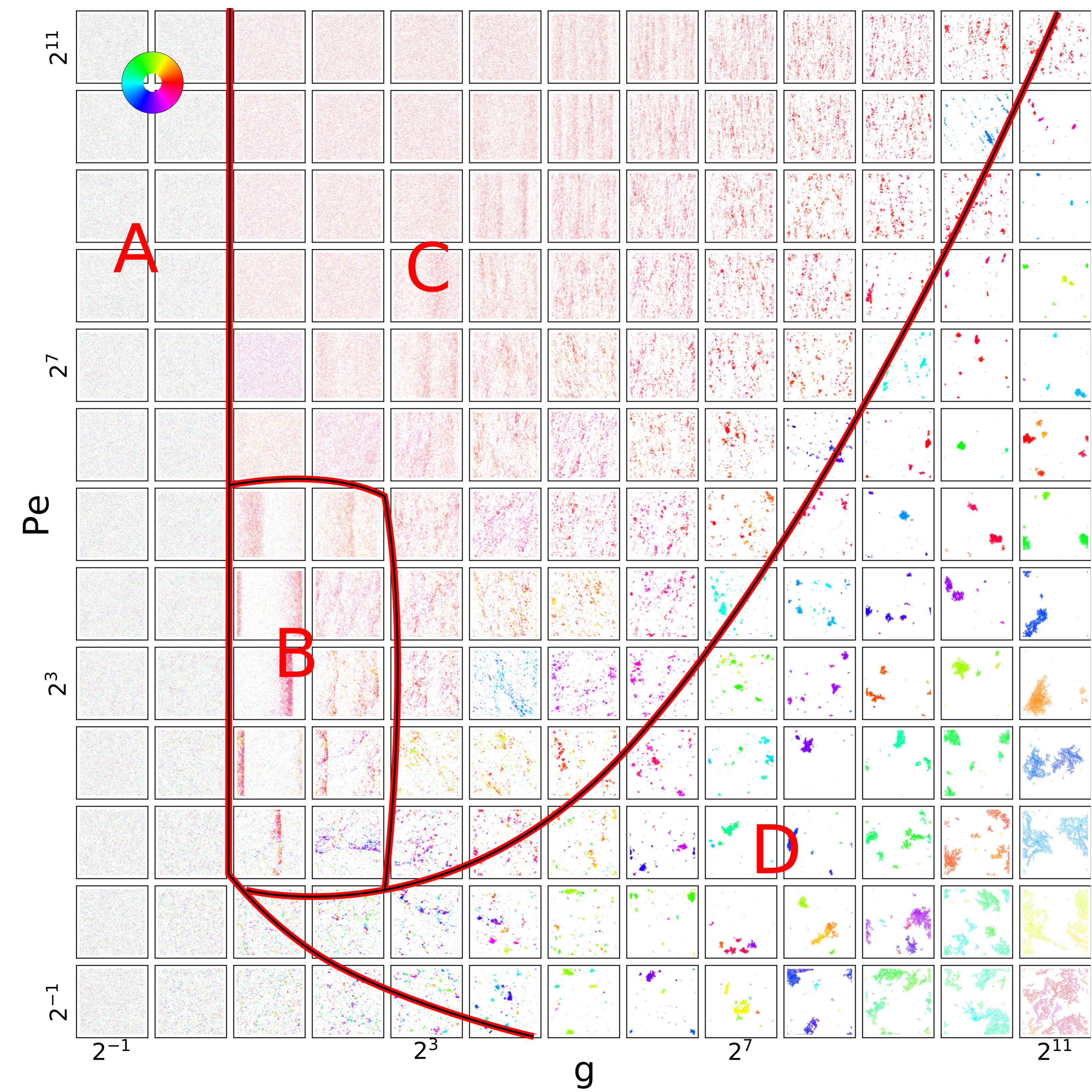}
    \caption{Phase diagram with representative snapshots of the Mean-Angle model without repulsion as a function of the coupling strength $\text{g}$ and Peclet number $\text{Pe}$. 
    Each particle is colored according to its heading direction (as shown in the top-left color disk). 
    The A-D labels identify the different regions discussed in the main text and the red lines sketch the boundaries between them.
    When comparing to the case with repulsion in Fig.~\ref{fig:ABC}, we observe similar features for low $\text{g}$ but significant differences for high $\text{g}$. 
    In particular, in this case without repulsion, we note that the clusters in region D have much higher density and that no homogeneous region E is found for high $\text{g}$ values.}
    \label{fig:snapshot-diagram-without-repulsion}
\end{figure*}

Figure \ref{fig:snapshot-diagram-without-repulsion} presents the phase diagram for the MA model with $\mu=0$.
The changes in this diagram with respect to the case with repulsion are similar to those observed for the MS and the SV models, but present significant differences when compared to the AS diagrams.
We will begin by analyzing this MA diagram, also representative of the MS and SV cases, and then discuss the AS case at the end of this section. 
In addition, we present in the Supplementary Material (Supp.Fig.\ref{fig:snapshot-diagram-sin-4-no-repulsion}) a phase diagram equivalent to Fig.~\ref{fig:snapshot-diagram-without-repulsion} for the AS model and snapshots equivalent to those in Fig.~\ref{fig:snap-shot-models-with-repulsion} but with $\mu=0$ for all models, where the effects of removing repulsive forces can be further compared.

We begin by noting that Fig.~\ref{fig:snapshot-diagram-without-repulsion} does not present significant changes for low $\text{g}$ values with respect to Fig.~\ref{fig:snapshot-diagram}.
In the disordered region A with $g \lesssim 2$, no clusters are formed due to the weak alignment forces and the spatial distribution remains homogeneous.
The structure also remains very similar in region B, since the density-order coupling mechanism that produces the bands is still present for $\mu=0$.
These bands appear to be thinner, however, as the lack of repulsion allows their local density to become higher.
We also note that region B extends to slightly lower values of $\text{Pe}$.
For example, the $(\text{Pe},g)=(2,2)$ snapshot corresponds to a disordered state in Fig.~\ref{fig:snapshot-diagram} but to an ordered state with a band in Fig.~\ref{fig:snapshot-diagram-without-repulsion}.
This appears to be because the very thin, high density band in the latter (which can only form without repulsive forces) helps the system remain ordered at higher noise levels.
Finally, as in the $\mu=1$ case, we find a low $\text{g}$ subregion of region C, above region B, where the homogeneity is due to finite-size effects.

For higher values of $\text{g}$, the diagrams display significant differences.
Indeed, the homogeneous ordered region C presents changes in, both, its states and its boundaries.
The clusters that form as $\text{g}$ is increased are now elongated perpendicular to the heading direction, instead of along it. 
In addition, the density fluctuations now form ripples at high $\text{Pe}$ values that were not observed for $\mu=1$.
Finally, for $\text{Pe} \lesssim 1$ and high $\text{g}$, we now find ordered states.
This appears to be due to the high density clusters that form without repulsion, since these remain aligned over longer distances and thus overcome the short persistence length of individual particles.

The boundary between regions C and D cannot be defined in Fig.~\ref{fig:snapshot-diagram-without-repulsion} the same way as it was in Fig.~\ref{fig:ABC}, because the clusters that now form in region D never span large spatial scales.
Indeed, due to the lack of repulsion, as we move into region D a few high density clusters start gathering most of the particles while collapsing into smaller areas.
Since our focus is not on determining the exact transition lines, we draw an approximate boundary between these regions when this is observed (which could be made more precisely by examining the distribution of cluster sizes \cite{martin2018collective}) .
As we move further into region D, increasing $\text{g}$ and decreasing $\text{Pe}$, the clusters gather more particles and become even denser, appearing to behave as rescaled self-propelled units that move collectively for long periods of time. 
Their dynamics become very slow, however, so we cannot be certain of their long-term behavior in the thermodynamic limit. 

As $\text{g}$ continues to be increased, no homogeneous region E appears in Fig.~\ref{fig:snapshot-diagram-without-repulsion}, in contrast to what was observed in Fig.~\ref{fig:ABC}, because agents now collapse into clusters due to the lack of repulsion.
As we approach the corner of the diagram with high $\text{g}$ and low $\text{Pe}$, however, we note that these clusters start to spread out due to a different mechanism.
In this regime, the aligning dynamics is so fast that the particles immediately start moving with almost the exact same velocity of any cluster with which they come into contact. This leads them to form extended, semirigid dendritic structures through a mechanism that is reminiscent of the Diffusion Limited Aggregation (DLA) process \cite{witten1981diffusion,paoluzzi2018fractal}.

Finally, we turn our attention to the phase diagram of the AS model without repulsion (included in the Supplementary Material). As mentioned earlier in this subsection, this is the only case that changes in ways that are significantly different from the other models when repulsive forces are removed.
Although the same A-D regions can be observed, we can highlight two important differences. First, the boundary between regions B, C and region D appears at much higher $\text{Pe}$ for the AS model than for other models without repulsion.
Second, the AS model displays relatively ordered states even for the lowest $\text{g}=2^{-1}$ coupling strength values in the diagram (at $\text{Pe} \approx 2$, for example).
We interpret both of these features as resulting from the additive nature of the AS model, which leads to a stronger density-alignment coupling that results in high density clusters. 
Indeed, these clusters will reduce the size of regions B and C, thus moving their boundary with region D upwards, as they prevent the formation of bands and homogeneous states. This is consistent with the results recently reported in \cite{chepizhko2021revisiting, kursten2021quantitative}, although we point out that we do not view the additive case as producing fundamentally different dynamics, but as instead displacing the boundaries in the phase space. 
The formation of high density clusters also explains the presence of ordered states even at very low $\text{g}$ and $\text{Pe}$ levels, since these can reach much higher persistence lengths than individual particles. 

\section{Discussion and conclusion}
\label{sec:conclusion}

%
We have considered four continuous-time models of collective motion with alignment interactions in their versions with and without repulsion. 
We explored their phase space as a function of two nondimensional parameters (Peclet number and interaction strength) that control the persistence length of noninteracting particles and the alignment rate of interacting particles, respectively.
We also identified three different homogeneous ordered states and compared their properties to previous numerical results and analytical predictions.

%
Our simulations show that all the models with repulsion present similar phase diagrams, with regions that display either disordered states or ordered states with different density distributions (homogeneous, with density bands, or with clusters elongated along or perpendicular to the heading direction).
%
%
The resulting phase diagrams highlight the fact that the states of the SPP models typically depend strongly on at least two independent control parameters.
This suggests that the single parameter scans performed in much of the literature by only changing the noise level can miss potential emergent states.

In this context, when comparing to previous studies of the Vicsek model (VM), it is important to point out that in the VM we can also control $\text{Pe}$ in isolation by changing the particle speed. However, since the alignment strength is not a free parameter in the VM (given that it is a discrete time algorithm with instantaneous alignment) there is no way to control $\mathrm{g}$ independently. 
On the other hand, changing the angular noise $\sigma^2_\theta$, the standard control parameter in the VM, while keeping all other parameters constant is equivalent to keeping the ratio of the Peclet number over the coupling strength constant, which corresponds to moving diagonally in the $\text{Pe}-\mathrm{g}$ phase diagram. 
%

We also note that if we vary the noise while fixing $\text{Pe}$, the system will visit different regions. If we fix $\text{Pe}$ at high values, the system will visit regions A, B, and C, but rarely region D and never region E. Instead, at intermediate $\text{Pe}$ values, the system will only visit regions A and D. Finally, at low $\text{Pe}$ and high $\mathrm{g}$ values, it will only visit regions A and E.
%

The phase diagrams without repulsion resemble the corresponding cases with repulsion for low values of the coupling strength $\text{g}$, but also display significant differences, in particular, for high $\text{g}$.
Indeed, when repulsion is present, we found a new homogeneous state E (for $\text{g}$ values beyond the clustering state D) that had not been previously reported in the literature. 
Without repulsion, the high $\text{g}$ states produce instead high density clusters in all our models.

Previous works suggested a fundamental difference between alignment models that are based on averaging or on the summation of angular differences, corresponding here to the MS or the AS model, respectively \cite{chepizhko2021revisiting,kursten2021quantitative}. These differences have been linked to a different density dependence of the alignment interaction. In our case, the AS model with repulsion exhibits a qualitatively similar phase diagram when compared to the other three models, likely due the repulsion effectively inhibiting large density fluctuations and high-density clusters.  
However, a detailed analysis of particular states such as the band regime discussed in \cite{chepizhko2021revisiting}, which is beyond the scope of this study, may reveal important differences. Furthermore, we note that, in line with these previous results, the phase diagram of the AS model without repulsion (included in the Supplementary Material) shows significant differences when compared to the corresponding diagrams for our averaging models (shown in Fig.~\ref{fig:snapshot-diagram-without-repulsion}).

The general features of all our phase diagrams can be understood in terms of the interplay between the orientational and the positional dynamics. The orientation of each particle can be viewed as an internal state that depends on the alignment interactions with other particles and on the noise. This state then affects the positions, which in turn determine which particles interact. When repulsive forces are present, they will affect the positions of interacting particles, which can also change which particles interact, but not their orientations.
This perspective can be expressed in terms of an adaptive network, as described in \cite{AlHuJSP2003,Huepe_2011,PhysRevE.94.022415}, where each node is associated to a state that corresponds to its orientation and the links between nodes represent the interactions, defined through a changing proximity network that depends on positions.

The language of adaptive networks can be used to describe the different regions of the phase diagram as a balance between the node state dynamics and the connectivity dynamics.
For high $\text{Pe}$ and low $\text{g}$, the connectivity changes much faster than the states and can thus be approximated by random mixing, so the system is well described by a mean-field theory. 
By contrast, for low $\text{Pe}$ and high $\text{g}$, the connectivity changes much slower than the node states and the system can be approximated by a fixed interaction network with alignment dynamics, similar to the XY model.
In this limit, slowly evolving spatial structures emerge from the feedback between the interaction topology and the angular dynamics. This explains the strong differences that are observed with and without repulsive forces, since these will affect this topology and therefore lead to the convergence to other states.
Between these two limits, the node dynamics and network evolution are entangled, resulting in the region of highest complexity. This is reflected in the emergence of clusters with a rich variety of spatial features and clustering dynamics.

%
Our results appear to show some universal characteristics of all SPP models with alignment interactions, although these could display changes in different parameter regimes.
For example, in order to limit our explorations, we only considered a specific mean density, given by $\pi N R^2/(4L^2)=0.3$.
Although we did not observe significant differences for small changes of this mean density in a limited set of exploratory simulations of the MA model that we performed for $\pi N R^2/(4L^2)\in \left[0.2,0.4\right]$ (not included in this paper), the phase diagram could present significant changes at very low or very high density values.
Furthermore, we also reduced the parameter space by considering the same interaction range for the repulsion and alignment terms, so some of the observed structures could change if these are set to be different. We note, however, that much of the phase space appears to be very similar even without repulsive interactions, which implies that they are probably also similar for any repulsion range smaller than the alignment range.
Despite all these considerations, further parameter explorations will be needed in future work to fully understand the universality or specificity of the presented results.

%
Our analysis of the three distinct homogeneous polarized states identified in the phase space showed that two of these (C-2 and E) match different aspects of the predictions of the Toner-Tu continuous hydrodynamic theory, while not matching others.
We noted that the differences could be attributed to a lack of complete homogeneity in the spatial distributions.
The third state (C-1), which seems to be fully homogeneous, produces results that strongly differ from the theoretical results, since the observed homogeneity appears to stem from finite-size effects. Because of this, the result match better a well-mixed mean-field system than the infinite homogeneous state described by the Toner-Tu theory.
We thus note that, even if here C-1 is the result of finite-size effects, it could be observed in well mixed large-scale models with weaker coupling between local density and alignment. 
This could be the case, for example, in topological models that consider interactions between a fixed number of nearest neighbors or in models where the interactions are restricted to the first neighbors in a Voronoi tessellation \cite{ginelli2010relevance}. 
Given that the features of state C-1 seem to be captured by a mean-field approach, it would be interesting to develop this type of theory in future work to predict its observed properties and exponents. We note, however, that this state still displays giant number fluctuations, so the usual homogeneity assumptions of a standard mean-field approximation may not be satisfied and extensions of this approach may be required.


Taken together, our results confirm that the Toner-Tu theory is generally applicable to the different homogeneous polarized states identified in our SPP models with alignment interactions and repulsion.
However, all these states show significant deviations with respect to the predicted exponents, because the homogeneous density assumption is not properly satisfied in our simulations.
Although it is possible that in larger simulations the systems reach a scale at which the coarse grained density distribution is homogeneous, we did not find any evidence of this in our results. 
In addition, we note that the implications of the presence of giant number fluctuations at all scales are still not well understood and that they could result in spatial inhomogeneities that also emerge at all scales. 
Further analytical studies and large scale simulations could help elucidate this issue.

Finally, our work shows that the alignment based SPP models display a plethora of different states in rich phase diagrams with a diversity of density structures that go beyond the homogeneous ordered states that have been the focus of much research.
In many cases, these could be more relevant for experimental systems and display universal model-independent features, as well as specific model-dependent ones, that require further investigations. 
Potential extensions of our work include considering larger scales; different repulsion forces and ranges; variable speeds; Voronoi, topological, or non additive interactions; and algorithms that produce collective motion without depending on explicit alignment terms
\cite{ginelli2010relevance,ferrante2013collective,mishra2012collective,chepizhko2021revisiting}. 

\vspace{0.5cm}

\section*{Acknowledgement}

We thank J. Toner and B. Mahault for their valuable comments and discussions. 
YZ is grateful for financial support from the China Scholarship Council (CSC). 
PR and YZ acknowledges funding by the Deutsche Forschungsgemeinschaft (DFG, German Research Foundation) through the Emmy Noether Programm - RO 4766/2-1 and and under Germany’s Excellence Strategy – EXC 2002/1 “Science of Intelligence” – project number 390523135.
This research was partially supported by the National Natural Science Foundation of China (NSFC) under Grant No. 61374165.  The work of CH was partially funded by CHuepe Labs Inc.

\bibliographystyle{unsrt}
\bibliography{references}

\begin{thebibliography}{10}

\bibitem{theraulaz2002spatial}
Guy Theraulaz, Eric Bonabeau, Stamatios~C Nicolis, Ricard~V Sol{\'e}, Vincent
  Fourcassi{\'e}, St{\'e}phane Blanco, Richard Fournier, Jean-Louis Joly, Pau
  Fern{\'a}ndez, Anne Grimal, et~al.
\newblock Spatial patterns in ant colonies.
\newblock {\em Proceedings of the National Academy of Sciences},
  99(15):9645--9649, 2002.

\bibitem{bazazi2011nutritional}
Sepideh Bazazi, Pawel Romanczuk, Sian Thomas, Lutz Schimansky-Geier, Joseph~J
  Hale, Gabriel~A Miller, Gregory~A Sword, Stephen~J Simpson, and Iain~D
  Couzin.
\newblock Nutritional state and collective motion: from individuals to mass
  migration.
\newblock {\em Proceedings of the Royal Society B: Biological Sciences},
  278(1704):356--363, 2011.

\bibitem{DeThViReview2012}
Andreas Deutsch, Guy Theraulaz, and Tam\'as Vicsek.
\newblock Collective motion in biological systems.
\newblock {\em Interface Focus}, 2:689, December 2012.

\bibitem{cavagna2010scale}
Andrea Cavagna, Alessio Cimarelli, Irene Giardina, Giorgio Parisi, Raffaele
  Santagati, Fabio Stefanini, and Massimiliano Viale.
\newblock Scale-free correlations in starling flocks.
\newblock {\em Proceedings of the National Academy of Sciences},
  107(26):11865--11870, 2010.

\bibitem{cavagna2015short}
Andrea Cavagna, Lorenzo Del~Castello, Supravat Dey, Irene Giardina, Stefania
  Melillo, Leonardo Parisi, and Massimiliano Viale.
\newblock Short-range interactions versus long-range correlations in bird
  flocks.
\newblock {\em Physical Review E}, 92(1):012705, 2015.

\bibitem{ballerini2008interaction}
Michele Ballerini, Nicola Cabibbo, Raphael Candelier, Andrea Cavagna, Evaristo
  Cisbani, Irene Giardina, Vivien Lecomte, Alberto Orlandi, Giorgio Parisi,
  Andrea Procaccini, et~al.
\newblock Interaction ruling animal collective behavior depends on topological
  rather than metric distance: Evidence from a field study.
\newblock {\em Proceedings of the national academy of sciences},
  105(4):1232--1237, 2008.

\bibitem{Breder_54}
C.M. Breder.
\newblock {Equations Descriptive of Fish Schools and Other Animal
  Aggregations}.
\newblock {\em Ecology}, 35(3):361--370, 1954.

\bibitem{couzin05}
I.~D. Couzin, J.~Krause, N.~R. Franks, and S.~A. Levin.
\newblock Effective leadership and decision-making in animal groups on the
  move.
\newblock {\em Nature}, 433:513--516, 2005.

\bibitem{westley2018collective}
Peter~AH Westley, Andrew~M Berdahl, Colin~J Torney, and Dora Biro.
\newblock Collective movement in ecology: from emerging technologies to
  conservation and management, 2018.

\bibitem{deseigne2010collective}
Julien Deseigne, Olivier Dauchot, and Hugues Chat{\'e}.
\newblock Collective motion of vibrated polar disks.
\newblock {\em Physical review letters}, 105(9):098001, 2010.

\bibitem{geyer2018sounds}
Delphine Geyer, Alexandre Morin, and Denis Bartolo.
\newblock Sounds and hydrodynamics of polar active fluids.
\newblock {\em Nature materials}, 17(9):789--793, 2018.

\bibitem{lavergne2019group}
Fran{\c{c}}ois~A Lavergne, Hugo Wendehenne, Tobias B{\"a}uerle, and Clemens
  Bechinger.
\newblock Group formation and cohesion of active particles with visual
  perception--dependent motility.
\newblock {\em Science}, 364(6435):70--74, 2019.

\bibitem{SumpterBook}
David J.~T. Sumpter.
\newblock {\em {Collective Animal Behavior}}.
\newblock {Princeton University Press}, 2010.

\bibitem{romanczuk2012active}
Pawel Romanczuk, Markus B{\"a}r, Werner Ebeling, Benjamin Lindner, and Lutz
  Schimansky-Geier.
\newblock Active brownian particles.
\newblock {\em The European Physical Journal Special Topics}, 202(1):1--162,
  2012.

\bibitem{VicsekZafeirisReview2012}
Tam\'as Vicsek and Anna Zafeiris.
\newblock Collective motion.
\newblock {\em Phys. Rep.}, 517(3–4):71 -- 140, 2012.

\bibitem{SwarmRoboticsReview2013}
Manuele Brambilla, Eliseo Ferrante, Mauro Birattari, and Marco Dorigo.
\newblock Swarm robotics: A review from the swarm engineering perspective.
\newblock {\em Swarm Intelligence}, accepted for publication.

\bibitem{vicsek1995novel}
Tam{\'a}s Vicsek, Andr{\'a}s Czir{\'o}k, Eshel Ben-Jacob, Inon Cohen, and Ofer
  Shochet.
\newblock Novel type of phase transition in a system of self-driven particles.
\newblock {\em Physical review letters}, 75(6):1226, 1995.

\bibitem{barberis2018emergence}
Lucas Barberis.
\newblock Emergence of a single cluster in vicsek's model at very low noise.
\newblock {\em Physical Review E}, 98(3):032607, 2018.

\bibitem{gregoire2004onset}
Guillaume Gr\'egoire and Hugues Chat\'e.
\newblock Onset of collective and cohesive motion.
\newblock {\em Phys. Rev. Lett.}, 92:025702, Jan 2004.

\bibitem{kursten2020dry}
R{\"u}diger K{\"u}rsten and Thomas Ihle.
\newblock Dry active matter exhibits a self-organized cross sea phase.
\newblock {\em Physical Review Letters}, 125(18):188003, 2020.

\bibitem{morin2015collective}
Alexandre Morin, Jean-Baptiste Caussin, Christophe Eloy, and Denis Bartolo.
\newblock Collective motion with anticipation: Flocking, spinning, and
  swarming.
\newblock {\em Physical Review E}, 91(1):012134, 2015.

\bibitem{martin2018collective}
Aitor Mart{\'\i}n-G{\'o}mez, Demian Levis, Albert D{\'\i}az-Guilera, and
  Ignacio Pagonabarraga.
\newblock Collective motion of active brownian particles with polar alignment.
\newblock {\em Soft matter}, 14(14):2610--2618, 2018.

\bibitem{czirok1996formation}
Andr{\'a}s Czir{\'o}k, Eshel Ben-Jacob, Inon Cohen, and Tam{\'a}s Vicsek.
\newblock Formation of complex bacterial colonies via self-generated vortices.
\newblock {\em Physical Review E}, 54(2):1791, 1996.

\bibitem{mishra2012collective}
Shradha Mishra, Kolbj{\o}rn Tunstr{\o}m, Iain~D Couzin, and Cristi{\'a}n Huepe.
\newblock Collective dynamics of self-propelled particles with variable speed.
\newblock {\em Physical Review E}, 86(1):011901, 2012.

\bibitem{toner2005hydrodynamics}
John Toner, Yuhai Tu, and Sriram Ramaswamy.
\newblock Hydrodynamics and phases of flocks.
\newblock {\em Annals of Physics}, 318(1):170--244, 2005.

\bibitem{chate2008collective}
Hugues Chat{\'e}, Francesco Ginelli, Guillaume Gr{\'e}goire, and Franck
  Raynaud.
\newblock Collective motion of self-propelled particles interacting without
  cohesion.
\newblock {\em Physical Review E}, 77(4):046113, 2008.

\bibitem{mahault2019quantitative}
Beno{\^\i}t Mahault, Francesco Ginelli, and Hugues Chat{\'e}.
\newblock Quantitative assessment of the toner and tu theory of polar flocks.
\newblock {\em Physical Review Letters}, 123(21):218001, 2019.

\bibitem{solon2015pattern}
Alexandre~P Solon, Jean-Baptiste Caussin, Denis Bartolo, Hugues Chat{\'e}, and
  Julien Tailleur.
\newblock Pattern formation in flocking models: A hydrodynamic description.
\newblock {\em Physical Review E}, 92(6):062111, 2015.

\bibitem{caussin2014emergent}
Jean-Baptiste Caussin, Alexandre Solon, Anton Peshkov, Hugues Chat{\'e},
  Thierry Dauxois, Julien Tailleur, Vincenzo Vitelli, and Denis Bartolo.
\newblock Emergent spatial structures in flocking models: a dynamical system
  insight.
\newblock {\em Physical review letters}, 112(14):148102, 2014.

\bibitem{sese2018velocity}
Elena Sese-Sansa, Ignacio Pagonabarraga, and Demian Levis.
\newblock Velocity alignment promotes motility-induced phase separation.
\newblock {\em EPL (Europhysics Letters)}, 124(3):30004, 2018.

\bibitem{kyriakopoulos2019clustering}
Nikos Kyriakopoulos, Hugues Chat{\'e}, and Francesco Ginelli.
\newblock Clustering and anisotropic correlated percolation in polar flocks.
\newblock {\em Physical Review E}, 100(2):022606, 2019.

\bibitem{HuepeAldanaPhysA2008}
Cristi\'an Huepe and Maximino Aldana.
\newblock New tools for characterizing swarming systems: A comparison of
  minimal models.
\newblock {\em Physica A: Statistical Mechanics and its Applications},
  387(12):2809--2822, 2008.

\bibitem{ToTuPRE98}
John Toner and Yuhai Tu.
\newblock Flocks, herds, and schools: A quantitative theory of flocking.
\newblock {\em Phys. Rev. E}, 58:4828--4858, Oct 1998.

\bibitem{toner2012reanalysis}
John Toner.
\newblock Reanalysis of the hydrodynamic theory of fluid, polar-ordered flocks.
\newblock {\em Physical Review E}, 86(3):031918, 2012.

\bibitem{bertin2009hydrodynamic}
Eric Bertin, Michel Droz, and Guillaume Gr{\'e}goire.
\newblock Hydrodynamic equations for self-propelled particles: microscopic
  derivation and stability analysis.
\newblock {\em Journal of Physics A: Mathematical and Theoretical},
  42(44):445001, 2009.

\bibitem{ihle2011kinetic}
Thomas Ihle.
\newblock Kinetic theory of flocking: Derivation of hydrodynamic equations.
\newblock {\em Physical Review E}, 83(3):030901, 2011.

\bibitem{grossmann2013self}
Robert Grossmann, Lutz Schimansky-Geier, and Pawel Romanczuk.
\newblock Self-propelled particles with selective attraction--repulsion
  interaction: from microscopic dynamics to coarse-grained theories.
\newblock {\em New Journal of Physics}, 15(8):085014, 2013.

\bibitem{peshkov2014boltzmann}
Anton Peshkov, Eric Bertin, Francesco Ginelli, and Hugues Chat{\'e}.
\newblock Boltzmann-ginzburg-landau approach for continuous descriptions of
  generic vicsek-like models.
\newblock {\em The European Physical Journal Special Topics},
  223(7):1315--1344, 2014.

\bibitem{grossmann2012active}
Robert Grossmann, Lutz Schimansky-Geier, and Pawel Romanczuk.
\newblock Active brownian particles with velocity-alignment and active
  fluctuations.
\newblock {\em New Journal of Physics}, 14(7):073033, 2012.

\bibitem{farrell2012pattern}
FDC Farrell, M~Cristina Marchetti, D~Marenduzzo, and J~Tailleur.
\newblock Pattern formation in self-propelled particles with density-dependent
  motility.
\newblock {\em Physical review letters}, 108(24):248101, 2012.

\bibitem{chepizhko2021revisiting}
Oleksandr Chepizhko, David Saintillan, and Fernando Peruani.
\newblock Revisiting the emergence of order in active matter.
\newblock {\em Soft Matter}, 2021.

\bibitem{kursten2021quantitative}
R{\"u}diger K{\"u}rsten and Thomas Ihle.
\newblock A quantitative kinetic theory of flocking with
  three-particle-closure.
\newblock {\em arXiv preprint arXiv:2102.13231}, 2021.

\bibitem{ihle2013invasion}
Thomas Ihle.
\newblock Invasion-wave-induced first-order phase transition in systems of
  active particles.
\newblock {\em Physical Review E}, 88(4):040303, 2013.

\bibitem{peruani2010cluster}
Fernando Peruani, Lutz Schimansky-Geier, and Markus Baer.
\newblock Cluster dynamics and cluster size distributions in systems of
  self-propelled particles.
\newblock {\em The European Physical Journal Special Topics}, 191(1):173--185,
  2010.

\bibitem{tu1998sound}
Yuhai Tu, John Toner, and Markus Ulm.
\newblock Sound waves and the absence of galilean invariance in flocks.
\newblock {\em Physical review letters}, 80(21):4819, 1998.

\bibitem{giavazzi2017giant}
Fabio Giavazzi, Chiara Malinverno, Salvatore Corallino, Francesco Ginelli,
  Giorgio Scita, and Roberto Cerbino.
\newblock Giant fluctuations and structural effects in a flocking epithelium.
\newblock {\em Journal of Physics D: Applied Physics}, 50(38):384003, 2017.

\bibitem{ginelli2016physics}
Francesco Ginelli.
\newblock The physics of the vicsek model.
\newblock {\em The European Physical Journal Special Topics},
  225(11-12):2099--2117, 2016.

\bibitem{witten1981diffusion}
TA~Witten~Jr and Leonard~M Sander.
\newblock Diffusion-limited aggregation, a kinetic critical phenomenon.
\newblock {\em Physical review letters}, 47(19):1400, 1981.

\bibitem{paoluzzi2018fractal}
Matteo Paoluzzi, Marco Leoni, and M~Cristina Marchetti.
\newblock Fractal aggregation of active particles.
\newblock {\em Physical Review E}, 98(5):052603, 2018.

\bibitem{AlHuJSP2003}
Maximino Aldana and Cristi\'an Huepe.
\newblock Phase transitions in self-driven many-particle systems and related
  non-equilibrium models: A network approach.
\newblock {\em J. Stat. Phys.}, 112:135--153, 2003.

\bibitem{Huepe_2011}
Cristi{\'{a}}n Huepe, Gerd Zschaler, Anne-Ly Do, and Thilo Gross.
\newblock Adaptive-network models of swarm dynamics.
\newblock {\em New Journal of Physics}, 13(7):073022, jul 2011.

\bibitem{PhysRevE.94.022415}
Li~Chen, Cristi\'an Huepe, and Thilo Gross.
\newblock Adaptive network models of collective decision making in swarming
  systems.
\newblock {\em Phys. Rev. E}, 94:022415, Aug 2016.

\bibitem{ginelli2010relevance}
Francesco Ginelli and Hugues Chat{\'e}.
\newblock Relevance of metric-free interactions in flocking phenomena.
\newblock {\em Physical Review Letters}, 105(16):168103, 2010.

\bibitem{ferrante2013collective}
Eliseo Ferrante, Ali~Emre Turgut, Marco Dorigo, and Cristian Huepe.
\newblock Collective motion dynamics of active solids and active crystals.
\newblock {\em New Journal of Physics}, 15(9):095011, 2013.

\end{thebibliography}

\end{document}


\title{Supplementary Information for \\ ``Phases and homogeneous ordered states in alignment-based self-propelled particle models'' \\ \small Zhao et al}

\appendix

\section{Numerical approaches}
In this section, we provide additional details on the computational and numerical aspects of our work.

\subsection{Density and velocity field}
We use coarse-grained density and velocity fields to analyze the different states, e.g. to calculate different correlation function.
In each frame, we have the locations of particles $\left\{\vec{r}_i\right\}$ and their orientations $\left\{\theta_i\right\}$. We subdivide the $L \times L$ space into $N\times N$ boxes $\left\{B_{\vec{r}}\right\}$, the density in a box is represented by particle number divided by the area of a box 
$$\rho(\vec{r}) = \frac{N_B(\vec{r})}{L^2/N^2}.$$ 
The coarse-grained velocity is defined by the average velocity of the particles in the box
$$\vec{v}(\vec{r}) = \frac{\sum_{i\in B(\vec{r})}\vec{v}_i}{N_B(\vec{r})}.$$ 
If no particle in the box, the velocity is zero. Here $B(\vec{r})$ is the set of particles in the box $B_{\vec{r}}$, $N_B=\left|B(\vec{r})\right|$, $\vec{r}$ is the discretized location of the box.

\subsection{Correlation function in Fourier space}

With the velocity field $\vec{v}(\vec{r},t)$, we define the average velocity $\vec{V}(t)=\left\langle \vec{v}(\vec{r},t) \right\rangle$ and the velocity fluctuation field $\delta \vec{v}(\vec{r},t)=\vec{v}(\vec{r},t)-\vec{V}(t)$. Hence, $\tilde{\delta \vec{v}}(\vec{q},t)=\int_{\vec{r}}\delta\vec{v}(\vec{r},t)e^{-2\pi i\vec{q}\cdot\vec{r}}d\vec{r}$.

For the equal-time spatial correlation function in Fourier space (as Eq \ref{eq:fourier-spatial-correlation-function}), we calculate
\begin{equation}
    \label{eq:fourier-spatial-correlation-function-velocity-2}
    C(\vec{q}) = \langle |\tilde{\delta\vec{v}}(\vec{q},t) |^2 \rangle
\end{equation}
In our calculation, we used $L\approx 1830, \,\,  N=800$.


We use the same approach for Fourier transfer of the density correlation function, 
\begin{equation}
    \label{eq:fourier-spatial-correlation-function-density-2}
    S_{\rho}(\vec{q}) = \langle |\tilde{\rho}(\vec{q},t) |^2 \rangle
\end{equation}
where $\tilde{\rho}(\vec{q},t)=\int_{\vec{r}}\rho(\vec{r},t)e^{-2\pi i \vec{q}\cdot\vec{r}}d\vec{r}$, except applying a higher resolution (N=8000), in order to obtain information about the relative positions of particles at the microscopic scale.

\subsection{Density-order coupling}
For the density-order coupling, we use K-nearest-neighbor approach for the local density, and quantify local order through the polarization of the $K$ nearest neighbors. The qualitative result is independent of the specific choice of $K_{nn}>2$, and we used $K_{\mathrm{nn}} = 5$ in our analysis.

We randomly select $\left\{\vec{r}_i\right\}$ in the space, for each of them, $\vec{r}_i$, we find its $K_{\mathrm{nn}}$ nearest neighbors $\left\{i_k\right\}$ ($k\in \left\{1,2,..., K_{\mathrm{nn}}\right\}$) among particles, with the distances $\left\{D_{i_k}\right\}$, so there are $K_{\mathrm{nn}}$ particles within the circle with radius $D_{i_k}$ and center $\vec{r}_i$, so the corresponding density is 
\begin{equation}
    \centering
    \rho(\vec{r}_i) = \frac{K_{\mathrm{nn}}}{\pi D_{i_{K_{\mathrm{nn}}}}^2}
\end{equation}
and the corresponding local order is
\begin{equation}
    \centering
    \phi(\vec{r}_i) = \frac{1}{K_{\mathrm{nn}} v_0}\left |\sum \vec{v}_{i_k}\right |_{k\in \left\{1,2,..., K_{\mathrm{nn}}\right\}}
\end{equation}

For each state we select 1000 frames and 1000 random points for each frame. For all pairs of $\left\{\rho(\vec{r}_i),\phi(\vec{r}_i)\right\}$, we average $\phi(\vec{r}_i)$ on $\rho(\vec{r}_i)$'s dimension with bin method.

\section{The growth of cluster with increasing system size}

We investigated the clustering states of model a-a (Eq .~\ref{eq:model-average-angle}) with different group sizes.
For generality, we choose 3 points in the parameter space (respectively (Pe,g) = (32, 256), (128, 256), (8, 64)). We consider all cluster that consist of more than 50 particles. For each cluster, we measure the length of its projection along and perpendicular to the cluster's mean direction, correspondingly referred to as the "length" and "width" of the cluster.

\begin{figure}[ht]
    \centering
    \captionsetup{justification=raggedright}
    \includegraphics[width=0.9\columnwidth]{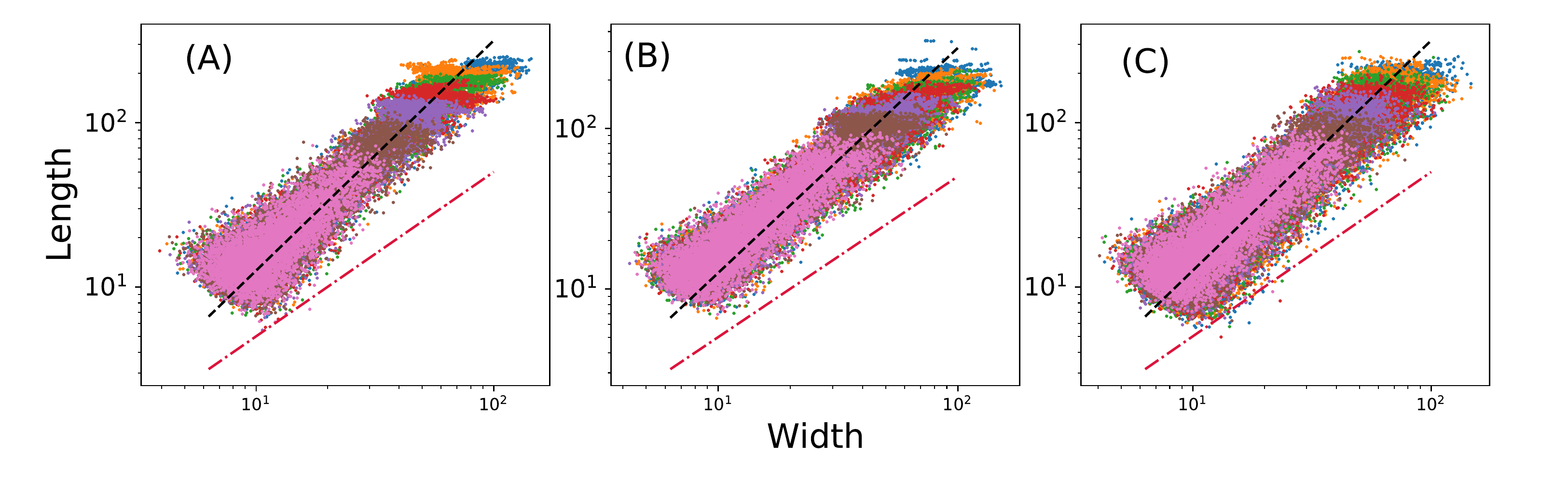}
    \includegraphics[width=0.9\columnwidth]{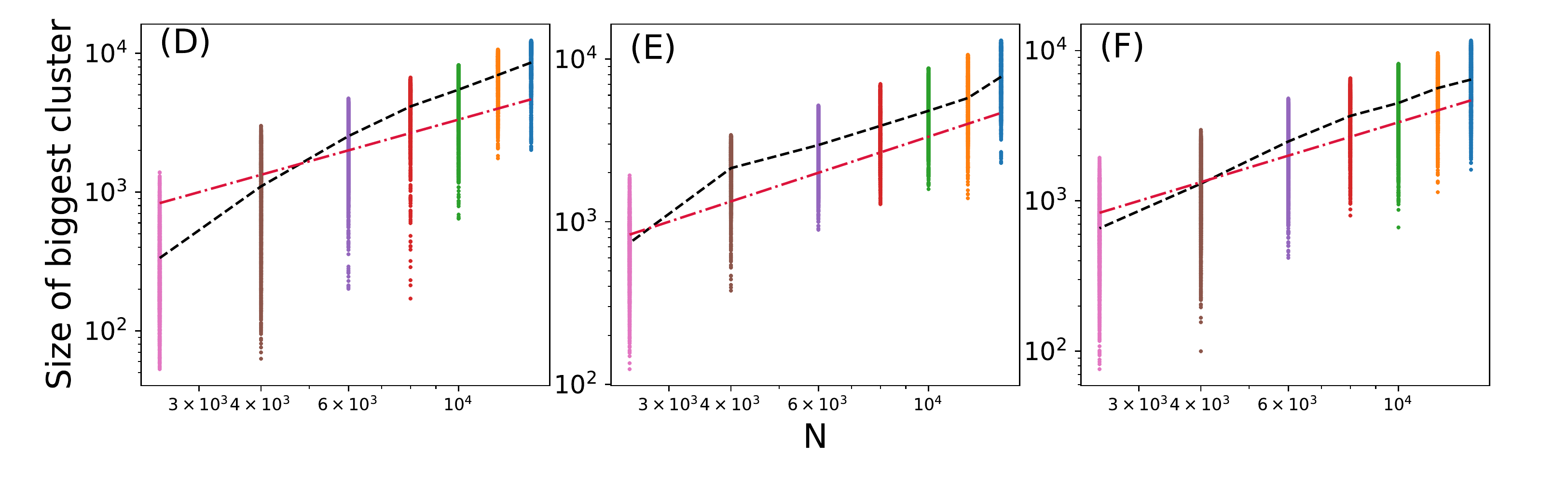}
    \includegraphics[width=0.9\columnwidth]{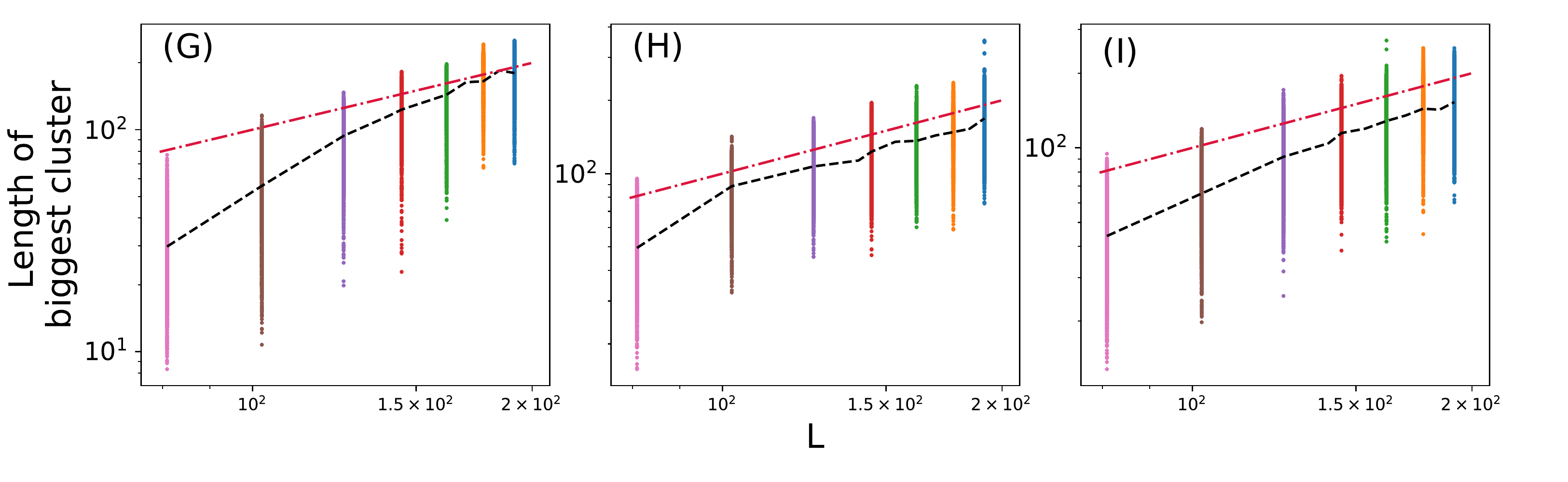}
    
    \caption{The correlations between: cluster width and cluster length (A,B,C), biggest cluster size and system size N (D,E,F), longest cluster length and system length L (G,H,I). Different color represents clusters collected from system of different system size: \textcolor{CarnationPink}{pink: 2500}, \textcolor{Brown}{brown: 4000}, \textcolor{Fuchsia}{purple: 6000}, \textcolor{red}{red: 8000}, \textcolor{green}{green: 10000}, \textcolor{orange}{orange: 12000}, \textcolor{blue}{blue: 14000}. From left to right respectively corresponds to (Pe,g) = (32, 256), (128, 256), (8, 64). In (A) to (C), each point represents the width and length of a cluster, black dash lines are linear fittings of the points, the red dot-dash lines represent the line $y=x$. In the second row, each point represents the size of the biggest cluster at one frame, with corresponding system size N, the black dash lines represent the average, the red dot-dash lines represent the line $y=0.3x$. In the third row, each point represents the length of the biggest cluster at on frame, with corresponding system length L, the black dash lines represent the average, the red dot-dash lines represent the line $y=x$.}
    \label{fig:clustering-width-length}
\end{figure}

We find, as shown in Fig.~\ref{fig:clustering-width-length} that the length and width of a cluster in a clustering state follow the power-law relationship, and the length grows faster than the linear system dimension $L$. This suggests that for sufficiently large systems with periodic boundary conditions, a single elongated cluster spanning the system should form. 

\section{The list of snapshots of alignment models without repulsion}

We post here the snapshot diagram for the models without repulsive force, as Fig.~\ref{fig:snap-shot-models-without-repulsion}:

\begin{figure}[ht]
\centering
\captionsetup{justification=raggedright}
\includegraphics[width=0.9\columnwidth]{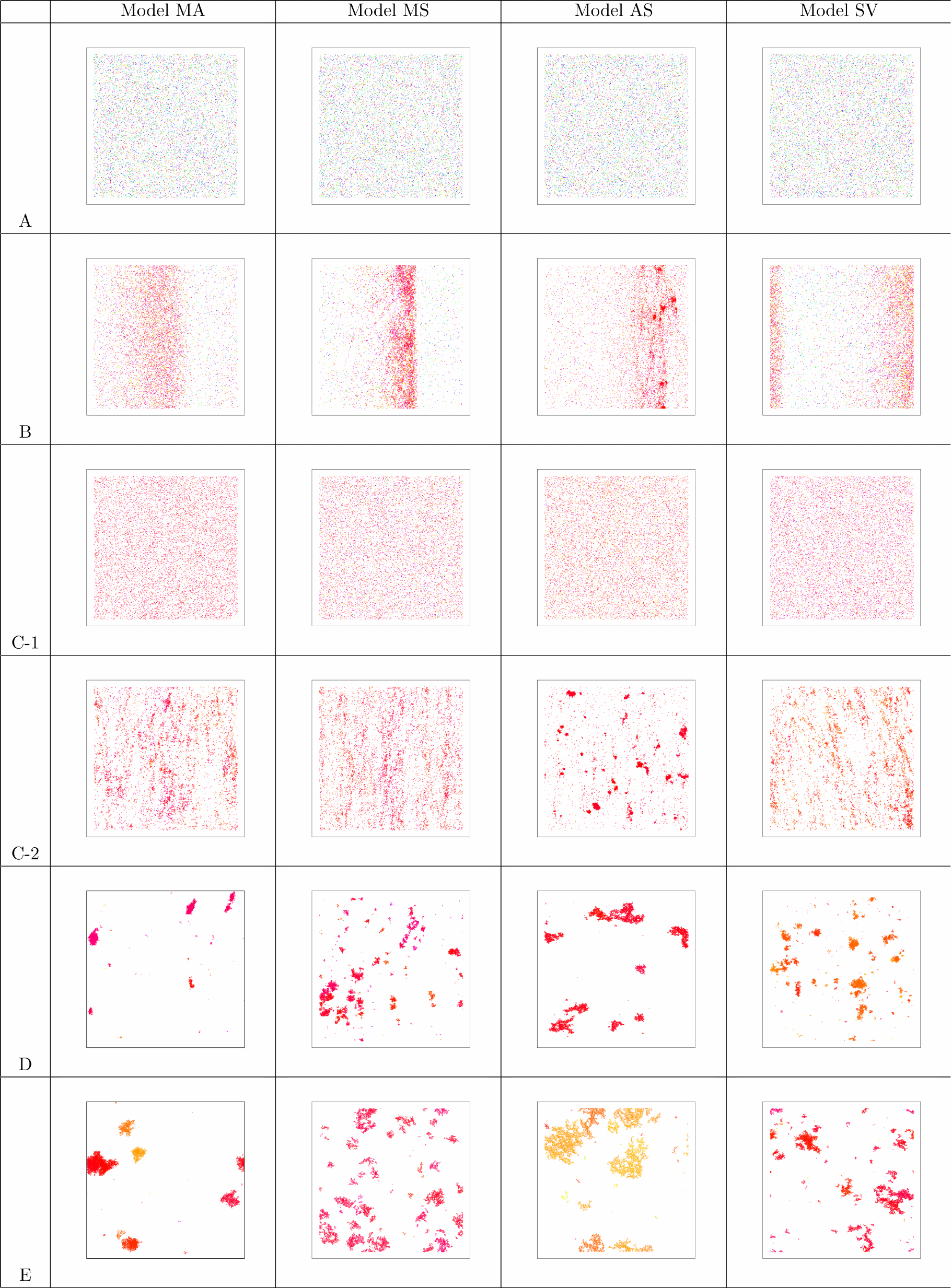}
\caption{The snapshots of states as in Fig.7 in the main text but in the absence of repulsion.}
\label{fig:snap-shot-models-without-repulsion}
\end{figure}

\section{The phase diagram of model AS without repulsion}

We post here the phase diagram model AS without repulsive force as Fig.~\ref{fig:snapshot-diagram-sin-4-no-repulsion}:

\begin{figure}[ht]
\centering
\captionsetup{justification=raggedright}
\includegraphics[width=1.0\columnwidth]{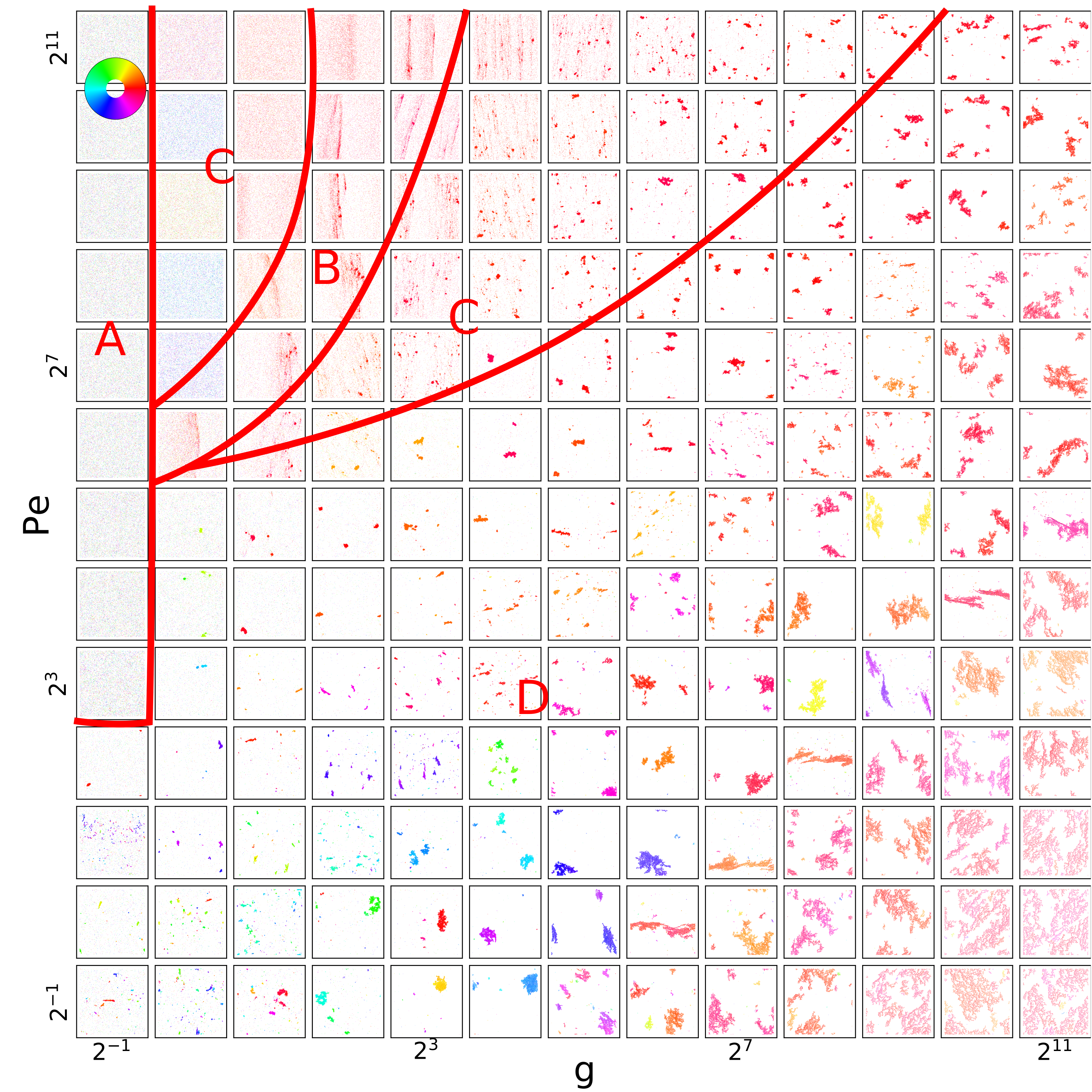}
\caption{The phase diagram of model AS without repulsive force.}
\label{fig:snapshot-diagram-sin-4-no-repulsion}
\end{figure}

